\documentclass[10pt,twocolumns]{IEEEtran}

\usepackage[english]{babel}
\usepackage[pdftex]{graphicx}
\usepackage{amsthm,amssymb,amsmath,subcaption,mathtools,epstopdf}
\usepackage[noadjust]{cite}
\usepackage{floatrow}
\usepackage{supertabular,booktabs}
\usepackage{multirow}
\usepackage[hyphens]{url}
\makeatletter

\newcommand{\Rmnum}[1]{\expandafter\@slowromancap\romannumeral #1@}
\makeatother

\renewcommand{\thefootnote}{\fnsymbol{footnote}}
\DeclareSymbolFont{boldoperators}{OT1}{cmr}{bx}{n}
\SetSymbolFont{boldoperators}{bold}{OT1}{cmr}{bx}{n}

\begin{document}

\renewcommand{\thefootnote}{\arabic{footnote}}
%\title{ Communication-Constrained Autonomous AV Routing and Control}
\title{\LARGE Communication-Constrained Routing and Traffic Control \\ for Autonomous Vehicles 
\thanks{G. Liu, S. Salehi, C.-C. Shen, and L. J. Cimini are with the University of Delaware, Newark, DE (e-mail: guangyi,\,salehi,\,cshen,\,cimini@udel.edu). E. Bala is with InterDigital (e-mail: Erdem.Bala@interdigital.com).}}
\author{Guangyi Liu, \textit{Student Member, IEEE}, Seyedmohammad Salehi, \textit{Student Member, IEEE}, \\  
Erdem Bala, \textit{Member, IEEE}, Chien-Chung~Shen, \IEEEmembership{Member,~IEEE}, 
and  
Leonard J. Cimini, \textit{Fellow, IEEE}
}
\maketitle

\begin{abstract}
Autonomous vehicles (AV) is an advanced technology that can bring convenience, improve the road-network throughput, and reduce traffic accidents. To enable higher levels of automation (LoA), massive amounts of sensory data need to be uploaded to the network for processing, and then, maneuvering decisions must be returned to the AV. Furthermore, 
%freed from the obligation of driving, 
passengers might have a higher transmission rate demands for various data hungry and delay-sensitive applications. However, due to frequent channel variations and imperfect cell deployments, guaranteeing a minimum transmission rate is impossible during a trip. In this work, the communication constraints of AVs are discussed. With these constraints, we present high-level concepts for the communication-constrained routing and traffic control optimization problems. First, to satisfy the minimum transmission rate requirements for the AV's LoA and certain applications used in the vehicle, we propose to divide the road network into two layers and perform a two-layered routing scheme; compared to some greedy methods, this scheme achieves a better balance between trip duration and communication coverage.
%Compared with Dijkstra and some greedy methods, the proposed two-layered routing scheme achieves a better balance between the trip duration and communication coverage. 
%Second, for video streaming  in the vehicle, we show that the deployment of mmWave base stations could provide better routing options in certain scenarios. Moreover, having routing information at base stations can facilitate the realization of mmWave vehicular communication.
Furthermore, for optimal traffic control purposes, we propose a key performance index (KPI)  to evaluate the traffic control capability of cellular systems. Then, two lemmas are proposed and proved to guide the goal of achieving optimal traffic flow with constrained communication resources. 
%To the best of our knowledge, this is the fist work in communication-constrained routing and traffic control for AVs.

\end{abstract}

\begin{IEEEkeywords}
autonomous vehicles, routing,  communication-constrained, traffic control, road-network throughput
\end{IEEEkeywords}

\section{Introduction}\label{introduction}

The technology of autonomous vehicles (AVs) has the potential to drastically reduce the energy consumption, traffic congestion, and collisions of vehicles
%, and thus can be a paradigm shift for the industry and for the society 
\cite{Bagloee2016,andrews2018}. In \cite{3gpp2018_2}, five levels of automation (LoA) of AVs are described: (0) no automation, (1) driver assistance, (2) partial automation, (3) conditional automation, (4) high automation, and (5) full automation.
With different LoA, AVs can enable new applications, such as dynamic ridesharing, platooning, remote driving, to name a few \cite{3gpp2018_2}. To achieve a higher LoA, AVs need to, throughout the entire trip, constantly communicate with neighboring AVs, as well as with traffic control services running in the infrastructure that continuously monitor the status of each AV and its surrounding environment to make driving decisions.
Specifically, because AVs might not have sufficient computing capacity for machine or deep learning based video/data processing, it has been proposed that portions of the data processing and driving decisions be delegated to the edge or to the cloud \cite{3gpp2018_2, tan2018}. Therefore, the availability, reliability, latency, and sustainable transmission rate of AV's connectivity with traffic control services running in the infrastructure are critical to achieve a high LoA \cite{3gpp2018_2,3gpp2018,shivaldova2013,stephens2013}. However, with AV's high mobility and the associated dynamic environment, these communication requirements bring significant challenges to the cellular infrastructure \cite{gozalvez2012,temel2016, 5Gwhitepaper2017}. 
%It is expected that eventually vehicles could be wholly autonomous \cite{Bagloee2016} without any human operations. 
%On the one hand, the routing and control of AVs require ultra-reliable low-latency wireless control links  On the other hand, 

%Also, freed from driving, passengers may have higher communication demands for all sorts of applications in their AVs, for example, virtual/augmented reality (VR/AR) applications may require a 50 to 200 Mbps transmission rate per passenger \cite{5Gwhitepaper2017,qualcomn2018}. %In contrast, in this paper, we focus on the joint routing and communication in situations when wireless network deployment is imperfect and communication is constraint \cite{gozalvez2012,temel2016}.  

%To bring good experience, future high throughput wireless network should be seamless and cover as many segments of the road as possible.

%For pedestrians and indoor scenarios, passengers can often be trapped in bad signal areas.

To provide wireless connectivity with higher quality-of-service (QoS) and reliability, fifth generation cellular systems (5G) wil support use cases such as enhanced mobile broadband (eMBB), massive machine type communications (mMTC), and ultra-reliable low-latency communications (URLLC) for mission-critical applications \cite{5Gwhitepaper2017,5G2018}. To enable these use cases, disruptive technologies, such as massive multiple-input and multiple-output (MIMO) and millimeter wave (mmWave), are expected to be deployed \cite{5Gwhitepaper2017}. 
%In particular, using hundreds to thousands of transmit antennas, massive MIMO lets a base station (BS) transmit to multiple users simultaneously to greatly increase the spectral efficiency \cite{marzetta2010,larsson2014}. %In addition, having orders of magnitude more bandwidth than what is currently available, mmWave communications are expected to achieve significantly higher throughputs \cite{marzetta2010}. 
In \cite{3gpp2018_3}, it is shown that four out of the nine 5G deployment scenarios support vehicular communication.  These are 1) Broadband Access in Dense Areas - providing 300 Mbps for vehicles with speeds up to 60 mph; 2) 50+ Mbps Everywhere - providing 50 Mbps with speeds up to 60 mph; 3) Mobile Broadband in Vehicles - providing 50 Mbps with speeds up to 300 mph; and 4) Ultra-Low Cost Broadband Access for Low Average Revenue Per User (ARPU) Areas - providing 10 Mbps with speeds up to 30 mph.
The actual deployment of 5G is likely to support a mixture of these scenarios across all regions.

Among the LoA, high and full automation would heavily rely on constant communication between an AV and the backend services running in the infrastructure. Typically, 50+ Mbps high-reliability low-latency communications is required during the entire trip to facilitate higher LoA and to accommodate any higher data rates demanded by passenger applications \cite{3gpp2018_2}. However, even when 5G is fully deployed, among the nine 5G deployment scenarios, only the ``Broadband Access in Dense Areas" use case can guarantee the sum transmission rate requirement imposed by the LoA of high and full automation. In addition, it could take several years to cover all the major areas with 5G, and the quality of communication is often worse around the cell boundaries. Therefore, without proper route planning, there is no guarantee that the sum-rate requirement imposed by the high LoA of the AVs can be sustained during the entire trip, especially when the density of AVs is high along certain road paths.  
 %a feasible plan for these 5G deployments might be difficult in high speed scenarios with rapidly time varying channels \cite{5Gwhitepaper2017,goldsmith2005}. Although, by testing and adjusting, network performance can be gradually improved, this process is costly and has limited room for improvement. With the proliferation of AV wireless applications, higher transmission rates and better reliability are required. Thus, the actual communication in certain areas might be far from ideal. %There may even be some signal out-of-coverage areas in certain places. %Also, it is unrealistic to deploy BS densely and let them perform coordinated multi-point (CoMP) because this large scale of CoMP is complex to realize. 

%For slow moving users and indoor scenarios, with continuous signal out-of-coverage areas, users can often be trapped with bad signal.In contrast, for 
%To increase AV LoA with the communication constraints described above, it is envisioned that the cooperation between vehicles and cellular infrastructures should be further deepened. On the one hand, to satisfy AVs' communication requirements,
In order to facilitate high LoA for the AVs during the entire
ride, here, we propose a communication-constrained routing (CCR)
problem, where, given the finite and different amounts of communication resources along the different road
segments (RSs) and intersections (RIs), the routes of AVs are deliberately selected,
such that the minimum sum transmission rate requirements of the
AVs can be satisfied at all times. In addition, we also propose a 
communication-constrained traffic control (CCTC) problem, where the 
communication resources available over different roads
can be utilized to maximize road-network throughput. To the best of our knowledge,
CCR and CCTC are new classes of routing and resource allocation
problems, respectively, that have not yet been explored. Only a recent 3GPP
document \cite{3gpp2018_2} elaborating the QoS aspects of automated driving
(Section 5.27.2.1) and remote driving (Section 5.28.2.2) describe selecting routes that can support the required communication QoS.

Previous work related to this topic includes energy-constrained routing 
and traffic optimization problems discussed in the context of electric vehicles
\cite{yu2018, wang2014, nejad2017}. However, the graph-based models adopted for road networks in electric vehicle routing cannot be properly adapted to either CCR or
CCTC. The difficulties can be attributed to the fact that (1) there are no detailed models connecting the key performance indexes (KPIs) of wireless
communication with those of AV movement; and (2) the edge weights of any routing graph computed from the sum transmission rates of wireless communication may
not be stable due to the fast time-varying nature of wireless channels. These two technical
issues will be specifically addressed in this paper.

%To the best of our knowledge, communication constrained AV routing and traffic control (CCR and CCTC) are, respectively, new classes of shortest path and resource allocation problems that have not yet been explored. 
%The electric vehicles energy-constrained routing and traffic optimization problems , adopting graph-based models for road-networks, were not adapted to CCR and CCTC. This is partially because 1) there is a lack of detailed modeling to connect communication key performance indexed (KPIs) and AV traffic KPIs; 2) the weights of the required routing graph computed by the transmission rates are not stable, with the fast variations of communication channels. These two points are to be addressed in this paper.%For 1), in this paper, we provide KPI connections by combining the most recent literature and technical reports in both AV and 5G areas. For 2), with massive MIMO BSs widely deployed in future cellular systems, small-scale fading is mitigated by channel hardening effect of massive MIMO \cite{marzetta2010}, such that the rates and the graph weights can be stable. %This indicates that, with 1) given, exploiting AV fast mobility, a supplementary approach to enhance AV communication can be applied: 

From the aspect of shortest path routing, one major challenge is to reduce the time complexity for large networks  \cite{rossi2018,li2009,yu2018,sanders2005, sanders2007}. In practice, a subset of critical nodes are often selected to divide the network into multiple layers, such that hierarchical routing can be applied with some preprocessing \cite{Bagloee2016,li2009,sanders2005,sanders2007}. However, the performance of hierarchical routing is strongly dependent on the selection of these critical nodes, which is often suboptimal. For CCR in large road networks, computing the theoretical minimum-trip-time route is a critical issue that is also addressed in this paper.

In terms of resource allocation, traffic control for AVs is an important issue that has been studied \cite{Milanes,boyles2018}. Without taking the communication constraints into consideration, backpressure-based traffic control, which maximizes the road-network throughput, is proposed in \cite{boyles2018}.
However, coupled with the CCR problem, when the density of AVs is high, the limited communication resources may not be able to satisfy the communication needs of all AVs. Under this constraint, we also study how to optimally utilize the available resources to maximize the road-network throughput.

 %How to realize joint traffic control and communication resources planning  Specifically, AV communication scheduling and resource allocation should be designed to .  %Specifically, global multi-task route planning functions in intelligent transportation should be  developed to 1) , 2) minimize the average trip time, 3) meet AV communication requirements. 

%Traditionally, navigation schemes like the one in the commonly-used Google Map compute routes which minimize the travel time using shortest path algorithms . The communication requirements from the AV passengers and the communication limitations caused by AV mobility, imperfect cellular infrastructures, geometry factors and available communication resources bring about new challenges to the routing problem. Like some other X-constrained routing problems \cite{yu2018, wang2014, nejad2017}, the decision to select shortest path route is limited by these challenges.

In this paper, we first describe an approach to modeling both the communication constraints for AVs
%in the context of massive MIMO 
and their effects on providing reliable wireless connectivity for higher LoA and reliable control. %With the deployment of massive MIMO BSs, small-scale fading can be mitigated by its channel hardening effect \cite{marzetta2010}, such that  the edge weights of the computed routing graph can be made stable. 
Then, by focusing on one individual AV, and without considering the impacts from other AVs, a non-convex optimization problem is formulated to minimize that AV's trip duration, subject to a transmission rate requirement. %To that end, using information about AV effective speed, that is operating below a specified vehicle speed, and the empirical rates collected under the ``intelligent transportation" initiative \cite{Boyang2017}, the weight on each edge of the graph can be computed. %Dijkstra's algorithm is a natural solution for this problem in small road-networks. However, due to its complexity, Dijkstra's algorithm cannot be applied to large road-networks \cite{li2009}. 
For faster routing computation in large road networks, a new hierarchical routing scheme is proposed, which divides the computed routing graph into a top layer containing all the base stations (BSs) and a bottom layer containing local graphs made up of RIs and RSs. The optimality of this hierarchical scheme is demonstrated. In addition, simulation results show that, compared with certain greedy algorithms, this two-layered routing scheme provides longer communication coverage durations and acceptable source-to-destination trip duration.

%Furthermore, for a specific application, video streaming, we show that a different communication constraint is required and the Dijkstra algorithm \cite{cormen2001} can be extended to tackle this challenge. It has been considered that there are many challenges for mmWave BSs for vehicular communication \cite{prelcic,va2016,wang2018}; however, considering video streaming as an example, we show that mmWave vehicular communication and routing with communication constraints can help each other's implementation. Also, with mmWave BSs, we show that joint AV traffic control and communication planning brings some new interesting problems.

For optimal traffic control, we focus on maximizing the road-network throughput subject to a constraint on the communication resources within a certain area. This is due to the fact that the complexity of performing city-wide, joint traffic control and route planning of all AVs is high. Therefore, we allow AVs to be navigated in a non-cooperative manner and focus on improving the road-network throughput for a given area.
Specifically, to characterize the traffic control performance of wireless communication for AV traffic control, we propose and investigate a new KPI for traffic control in cellular systems. Using this KPI, two lemmas are proved to determine the optimal speed of each AV and optimal frequency channel allocation across multiple cells. Numerical results are provided to show the performance gain.

The rest of the paper is organized as follows. In Section \ref{preliminary}, the impacts of communication on AV routing and traffic control are discussed, motivating the introduction of two concepts to facilitate the study of CCR and CCTC. 
Section \ref{main_1} studies the CCR problem, where a road network is modeled as a two-layer graph to facilitate a two-layered communication-constrained routing scheme. Simulation results are presented to demonstrate the effectiveness of the proposed scheme.
Section \ref{control} investigates the CCTC problem by first computing AV speed within a single cell to maximize road-network throughput. Then, to match traffic flow across adjacent cells, a spectrum re-balancing solution is introduced to maximize AV road-network throughput across multiple cells. The effectiveness of the solution is validated via numerical studies. Section \ref{last} concludes the paper and suggests future research directions.

\section{Impacts of Communication on Routing and Traffic Control AVs}\label{preliminary}

This section discusses the impacts of communication on the routing and traffic control of AVs, and introduces {\em effective speed map} (ESM) and {\em $\gamma$-rate cell} of a BS.
%In providing future wireless communication for AVs, there will be a difference between communication for data and communication for control. As shown in Fig. \ref{fig:overview}, part of the AVs require communication for data so that passengers can use applications like video streaming; meanwhile, all AVs require communication for control so that signaling can be received to guide their navigation. In 5G, the former corresponds to the concept of eMBB, and the latter corresponds to URLLC \cite{5Gwhitepaper2017}. In this section, we will describe the cellular technologies for both, and also the factors affecting them.
% Based on such impacts, we propose to construct the {\em effective speed map} (ESM) and the {\em empirical rate map} (ERM), in Subsections \ref{pre-control} and \ref{map_2}, to facilitate CCR and CCTC, respectively. %These two maps can be constructed by using the technique of big data analytics \cite{Boyang2017}. 
% Also, similar maps, like real-time High Definition mapping will be built for AV navigation \cite{5G2018}.   
% Similarly, \cite{5G2018} described the construction of high-definition (HD) maps needed for real-time situation awareness of road users.

%Also, with the initiatives of intelligent transportation and big data technology applied in the wireless industry \cite{Boyang2017}, real-time High Definition (HD) mapping is required for AV navigation (or routing) .
%The proposed ESM and ERM can easily update with cellular systems.  

\begin{figure}[htb]
	\centering
		\includegraphics[width=0.8\textwidth]{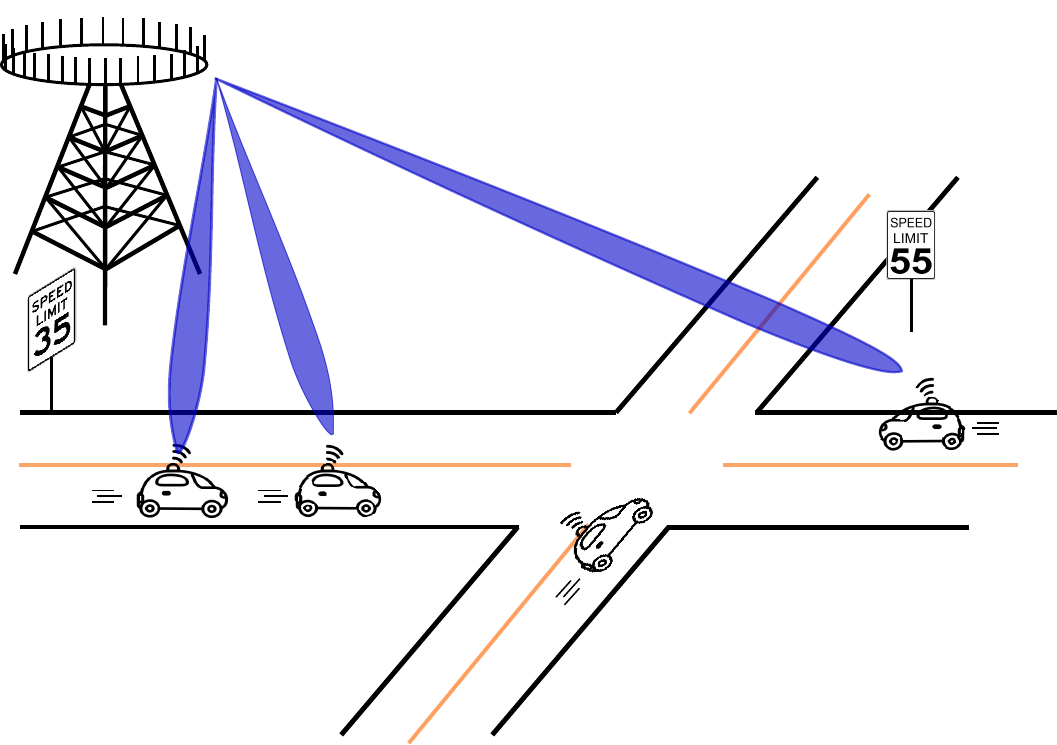}
		\caption{\label{fig:overview} Infrastructure to AV communication}
\end{figure}

\subsection{Effective Speed Map (ESM)}\label{pre-control}

The speed of the AVs is a critical parameter for both routing and traffic control. For routing, given the distance of a road path between the source and destination, the speed determines the driving time for an AV to complete its route. 
For traffic control, the throughput of the road network directly depends on the speed of the AVs.

To decide the speed of the AVs, the reliability, transmission rate, and latency of communication between AVs and the traffic control service running in the infrastructure are critical, in addition to the physical conditions of the roads, the density of vehicles, and the speed limit ($v_l$) imposed by the transportation authorities  \cite{stephens2013,3gpp2018_2,zachariou2011}.
% As discussed in \cite{3gpp2018_2}, to enable high LoA for AVs, certain minimum communication requirements of reliability, transmission rate, and latency, denoted by $\Gamma$ = [$\mathcal{R}_0, \mathcal{G}_0, \mathcal{A}_0$], must be satisfied for each AV\footnote{The higher the LoA, the more stringent the communication requirements $\Gamma$.}.
%From the perspective of control theory, \cite{ehrig2017} shows that, with the reliability of location-based control, location-based operations could be applied to ensure the safety of AV movements, including specifying the speed limit.
For instance, %it has been claimed in \cite{boyles2018} that, 
if the AVs have faster reaction times, they might be able to move faster and travel at closer distances with each other without compromising safety. %thereby effectively increasing the capacity of a road and reducing congestion 
This reaction time strongly depends on the reliability, transmission rate, and latency of the AV communication.% between AVs.% and the traffic control service running in the infrastructure.

In this paper, we propose an approach that uses the speed values, computed by maximizing the throughput of a road network (from Lemma 1 in Section \ref{control}), as the {\em effective speed} (values) of AVs for route planning (in Section \ref{main_1}).  Specifically, these speed values are associated with RSs of a road network to form its effective speed map (ESM) to be used by the routing function. For computing the trip duration and planning the routes, as with current vehicle navigation systems, we assume that the ESM does not change for the time period that a target AV is routed, and this AV travels exactly at this effective speed.

\subsection{$\gamma$-Rate Cell of a BS}\label{map_2}

%As mentioned in Section \ref{introduction}, passengers of AVs may use various high transmission rate applications. Due to the mobility of AVs, to reduce frequent BS hand-offs, it is more applicable to apply macro cell BSs to serve the AVs like eNB in LTE-A \cite{3gpp2016}.%, in LTE-A, New Radio, or future massive MIMO macro cell BSs. 

%I am not sure if gamma rate cell region has been defined before
% In this paper, we consider BSs that employ the technology of massive MIMO \cite{marzetta2010,larsson2014}, which can provide high transmission rates for many users simultaneously.
In \cite{5G2018}, BSs would control the AVs.
For a BS, geometrically, we define its \emph{$\gamma$-rate cell} to be the largest area surrounding the BS such that %(1) the measured downlink transmission rate\footnote{AV applications usually have rate requirements for both uplink and downlink. However, due to the uplink-downlink duality \cite{vishwanath2003}, for a certain BS, its uplink transmission rate is closely related to its downlink transmission rate. So, for simplicity, in this paper, we consider only the downlink.} at any location within the area is higher than the threshold of $\gamma$ Mbps,
(1) the probability of having a downlink transmission rate\footnote{AV applications can also have rate requirements for uplink transmissions. However, the methodology for solving the uplink CCR problem is the same.
%However, due to the uplink-downlink duality \cite{vishwanath2003}, for a certain BS, its uplink transmission rate is closely related to its downlink transmission rate. So, 
So, for simplicity, in this paper we consider only the downlink. } more than $\gamma$ Mbps at any location within the area, is higher than a threshold of 1$-\epsilon$, where $\epsilon$ is a very small number, 
and (2) for any two locations on the roads within this area, there exists a road path within the same area connecting them. 
% sequence of RSs 
Thus, each $\gamma$-rate cell is one contiguous region. Based on this definition, with rates $\gamma_1<\gamma_2$, for a specific BS, its $\gamma_1$-rate cell contains its $\gamma_2$-rate cell. The two $\gamma$-rate cells of two adjacent BSs are connected if their overlapping area covers at least one common RI or one common RS.
For example, owing to the channel hardening effect of massive MIMO
\cite{marzetta2010}, small-scale fading is mitigated and the magnitude of variations of transmission rates is small, so that the boundaries of $\gamma$-rate cells are stable over time even when $\epsilon$ goes to 0.
Although it is not necessary to keep track of the exact boundary of a $\gamma$-rate cell, for AV navigation, it is good enough to keep track of all the RIs and RSs within the $\gamma$-rate cell of a BS. 

In order to compute the $\gamma$-rate cells for the BSs, the probability that the transmission rate is above a given threshold $\gamma$ should be recorded every few meters along the RSs. To obtain this probability corresponding to this distance, all the rate measurements within this interval should be reported through the BSs to the route planning function, together with the corresponding GPS locations.

%In order to compute the $\gamma$-rate cells for the BSs, AVs measure their experienced transmission rates (for instance, one measurement every 10 meters) and report these measurements together with the corresponding GPS locations through the BSs to the route planning function.

%Note that due to the mobility of AVs, the measured user channel is not always equal to the actual channel. For simplicity, we assume the CSI delay to be the channel measuring interval $\tau$. Assuming Clarke's channel model, given the speed of AVs, the theoretical transmission rates of massive MIMO could be computed using the approximation methods described in \cite{papa2017}. 

%In the next section, we study the routing of a single vehicle by ignoring the impacts of other vehicles. However,
%in Section \ref{control}, we relax this assumption and propose the concept of a traffic control network, embedded within the cellular infrastructures, to extend ESM for optimal AV traffic control, with constrained communication resources.

\section{Communication-Constrained Routing (CCR)}\label{main_1}

The CCR problem
% \footnote{Note that, if the required transmission rate is satisfied in the entire area, the formulated problem is not useful. But this is too strong an assumption, considering the cost for such dense BS deployment with fast channel variations and severe interferences. Also, even though there are technologies to improve the transmission rates and signal coverage, many AV applications are being created, such that the rate requirement $\gamma$ is becoming higher.} 
can be stated as follows: find the shortest-time path from the source location to the destination location where the given minimum transmission rate can be sustained along the entire path at all times. In this section, we focus on routing one AV in a non-cooperative (selfish) manner. The AV is fully autonomous % \cite{Bagloee2016} requiring no human operation. Furthermore, 
with collision avoidance, so that the time spent by this AV on waiting at RIs and yielding to pedestrians and other AVs is considered negligible
% \footnote{For instance, with the sense-plan-act capability introduced in \cite{}, pedestrians can be properly avoided by AVs without stopping and waiting.} 
\cite{stephens2013,Bagloee2016,boyles2018}.

A road network with deployed BSs over a given area, as depicted in Fig. \ref{twolayer}(a), can be modeled as a graph $G_{\text{road}}(\mathcal{V},\mathcal{E}, \mathcal{M})$ \cite{wang2014}, where node set $\mathcal{V}$ denotes the set of RIs, edge set $\mathcal{E}$ represents the set of RSs connecting adjacent RIs, and $\mathcal{M}$ is the set of BSs deployed in the given area. For BS $m\in\mathcal{M}$, $G_{\text{road}}^m$ denotes the graph representing the part of the road network located within $m$'s $\gamma$-rate cell, as depicted in Fig. \ref{twolayer}(b).
%As shown in Figs. \ref{shen1} and \ref{shen2}, an RS is kept in one local graph only when the entire RS is within the corresponding $\gamma$-rate cell. 
For both $G_{\text{road}}$ and $G_{\text{cell}}^m$, the weight of an edge (RS) is the travel time of an AV over this edge, which is computed as the length of this edge divided by the corresponding effective speed obtained from the ESM, which was discussed in Subection \ref{pre-control}. 

\subsection{Two-Layered AV Routing Scheme}

%Exhaustively search the transmission rates on the ERM for all edges of  $G(\mathcal{V},\mathcal{E})$; if the transmission rate at a certain location is smaller than $\gamma$, then cut the corresponding edge from $G(\mathcal{V},\mathcal{E})$. Denote the new graph to be $G'(\mathcal{V},\mathcal{E})$. The classic Dijkstra's Algorithm can be directly applied on $G'(\mathcal{V},\mathcal{E})$ to obtain the route solution, for small road networks. However, Dijkstra's Algorithm has a complexity $O(|\mathcal{E}|\log_2|\mathcal{V}|)$, which means it is not suitable for large road networks \cite{li2009}. 

In \cite{li2009}, a road network was also modeled by a graph similar to $G_{\text{road}}$ but without the component $\mathcal{M}$.
To reduce routing complexity, \cite{li2009} selected a subset of critical RIs to perform hierarchical routing. However, the performance of this approach strongly depends on the RIs selected and is usually suboptimal.

\begin{figure}[ht]    
\begin{subfigure}{1\textwidth}
\centering
\includegraphics[width=0.8\linewidth]{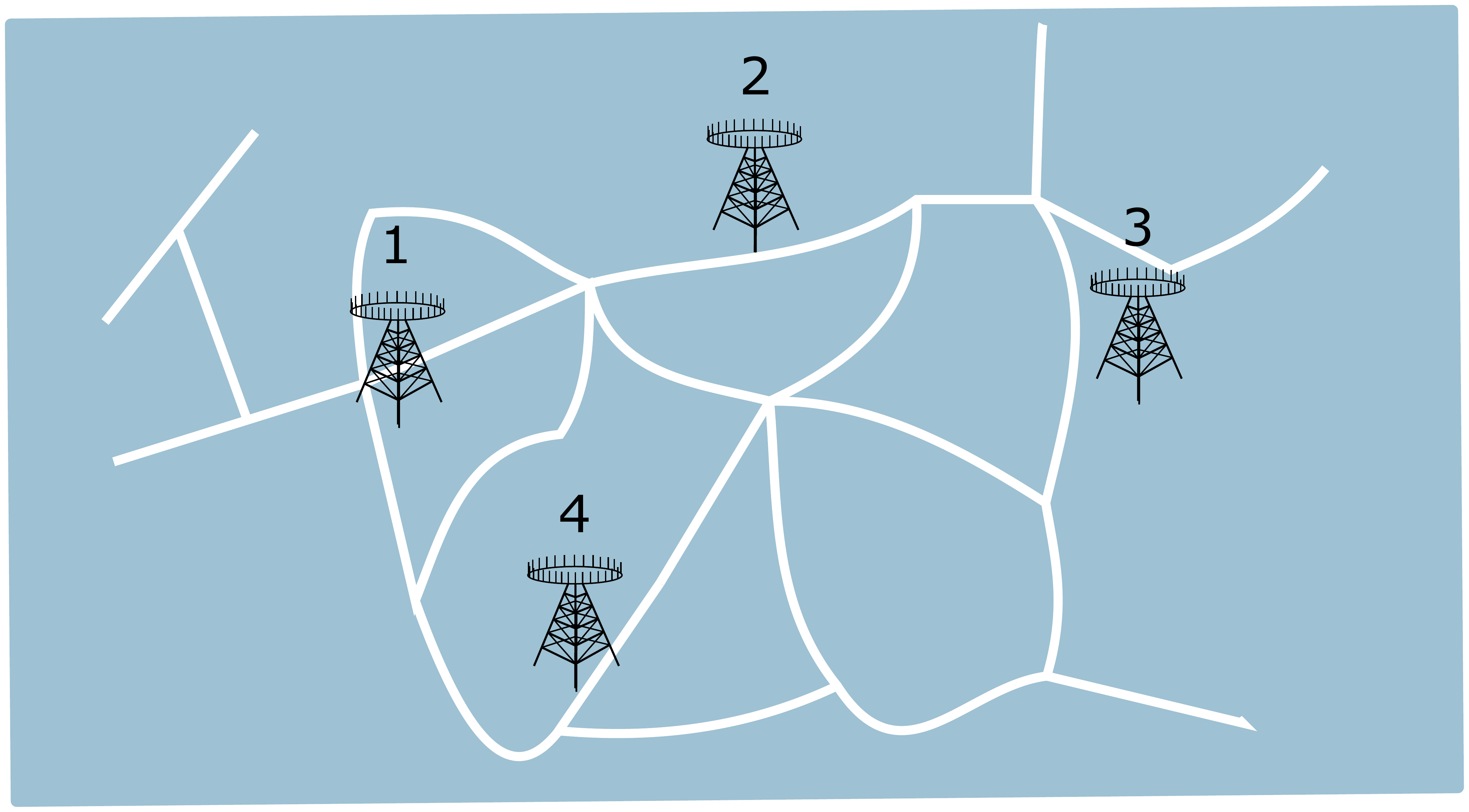}
 \caption{Road network $G_{\text{road}}$}       %\vspace{-80pt}
\label{shen1}
\end{subfigure}

\begin{subfigure}{1\textwidth}
\centering
\includegraphics[width=0.8\linewidth]{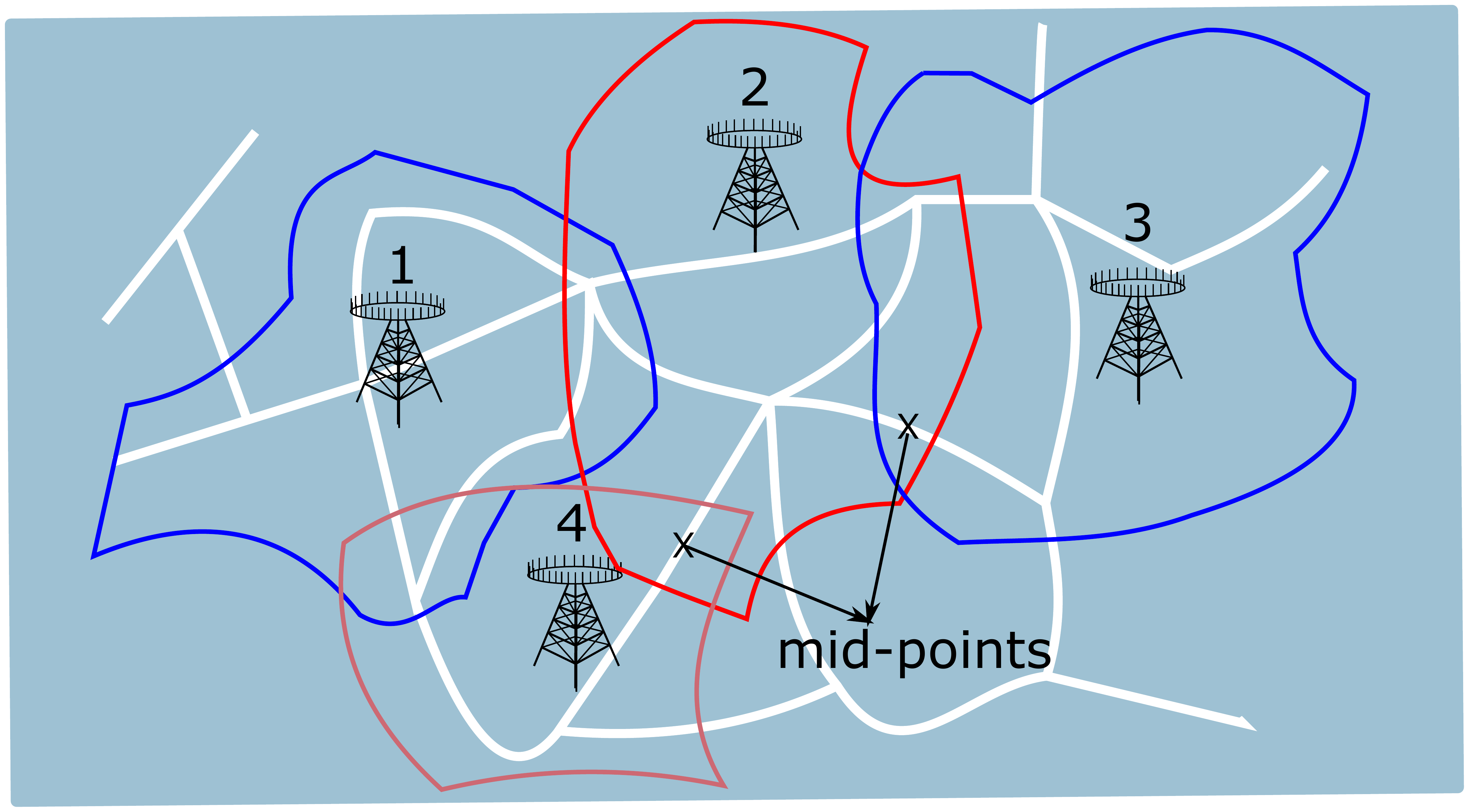}
\caption{$\gamma$-rate cells ($G_{\text{cell}}^1, G_{\text{cell}}^2, G_{\text{cell}}^3,~\text{and}~ G_{\text{cell}}^4$)}        \label{shen2}
\vspace{12pt}
\end{subfigure}
\begin{subfigure}{1\textwidth}
\centering
\includegraphics[width=0.8\linewidth]{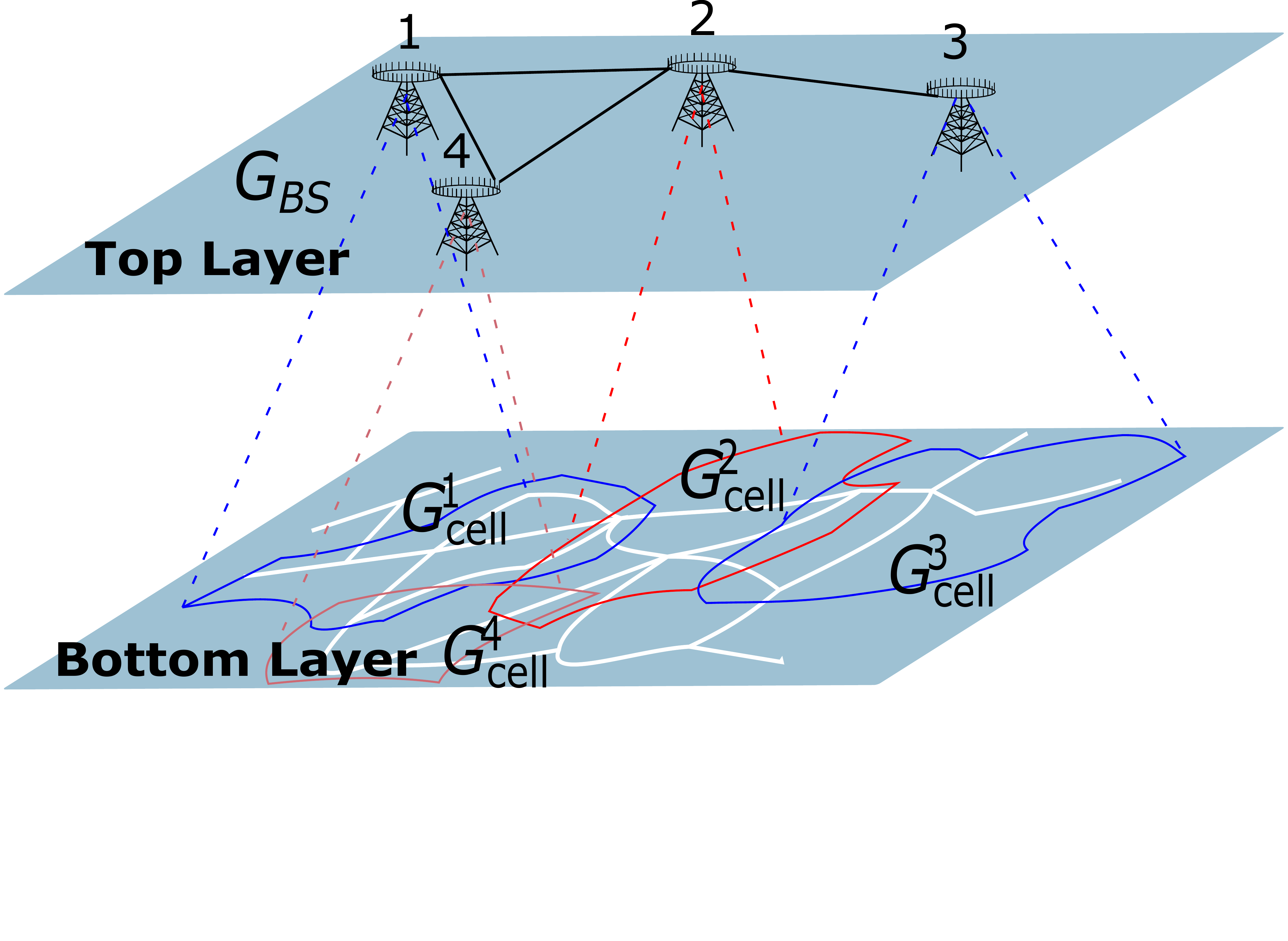}
\caption{Two-layer graph}        \label{shen3}
\end{subfigure}
\caption{Transfer a road network into two layers}
\label{twolayer}
\end{figure}

As in urban scenarios where the number of BSs is much smaller than the number of RIs (for instance, some cells have radii of as much as 1-2 km \cite{goldsmith2005}), in contrast to \cite{li2009}, we propose to 
divide $G_{\text{road}}(\mathcal{V},\mathcal{E}, \mathcal{M})$ into two layers to facilitate a two-layered routing scheme. The top layer is denoted by graph $G_{\text{BS}}$ representing the connectivity among $\gamma$-rate-cells, and the bottom layer consists of all the graphs of $\gamma$-rate-cells ($G_{\text{cell}}^m$) of these BSs (as shown in Fig. \ref{twolayer}(c)). Over these two layers,  {\em inter}-$\gamma$-rate-cell routing and {\em intra}-$\gamma$-rate-cell routing are conducted in sequence followed by a process of dynamic programming to compute the shortest path between the source and destination locations. The optimality of the proposed scheme is demonstrated at the end of this subsection. 
\begin{figure}[htb]
	\centering
	\includegraphics[width=0.8\textwidth]{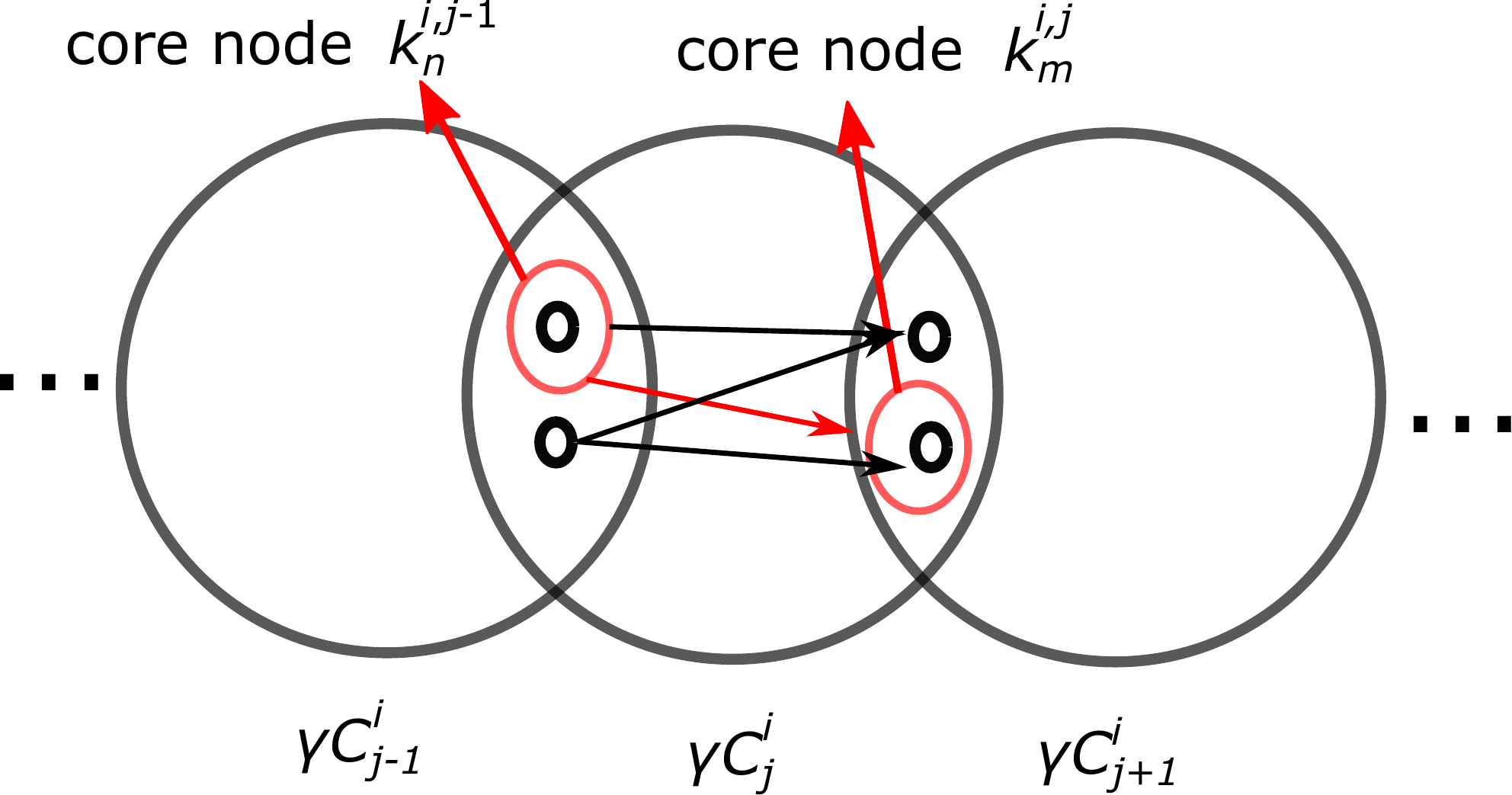}
	\caption{$\gamma$-rate cells and core nodes in path $P_i$. (The arrows in $\gamma C_{j}^i$ illustrate the proposed dynamic programming procedure.)}\label{fig: SD}
\end{figure}
%THIS CAPTION NEEDS TO BE CHANGED.  IT NEEDS TO BE MORE DESCRIPTIVE AN "SCHEMATIC DIAGRAM" SHOULD NOT BE IN TH E CAPTION.  SAME FOR NEXT FIGURE AND ALL FIGURES
Specifically, in $G_{\text{BS}}$,
% For top-level path selection in $G_{\text{BS}}$, 
there exists an edge between two $\gamma$-rate cells only when these two $\gamma$-rate cells are connected. 
% For simplicity, it is assumed that both the starting RI and the destination RI are covered by $\gamma$-rate cells; then these two BSs are the starting and destination BSs in $G_{\text{BS}}$. 
Given that two $\gamma$-rate cells may be connected in the road network, via multiple common RIs and/or RSs, the exact travel time (i.e., weight) over a (top-layer) edge in $G_{\text{BS}}$ between any two adjacent $\gamma$-rate cells is initially undefined. Therefore, for inter-$\gamma$-rate-cell routing, Breadth-First Search (BFS) is applied to find the set of all possible (top-layer) paths from the source $\gamma$-rate cell, through zero or more intermediate $\gamma$-rate cells, to the destination $\gamma$-rate cell\footnote{Source (destination) $\gamma$-rate cell is defined as the $\gamma$-rate cell that contains the source (destination) location of the AV.}. 
Let $\mathcal{P}=\{P_1, P_2, \cdot\cdot\cdot ,P_{|\mathcal{P}|}\}$ be the set of (top-layer) paths found, and $|P_i|$ be the number of $\gamma$-rate cells traveled in path $P_i$. For path $P_i$, [$\gamma C_{1}^i, \gamma C_{2}^i,\cdots, \gamma C_{|P_i|}^i$] denotes the sequence of $\gamma$-rate cells traveled (see the schematic diagram in Fig. \ref{fig: SD}) , where $\gamma C_{1}^i$ and $\gamma C_{|P_i|}^i$ are the source and destination $\gamma$-rate cells in $G_{\text{BS}}$, respectively. 
%
% So, the $\gamma$-rate cell of each BS $B_{j}^i$, except $B_{1}^i$ and $B_{|P_i|}^i$, must overlap with two other $\gamma$-rate cells: a former associated with $B_{j-1}^i$ and a latter associated with $B_{j+1}^i$. 

For path $P_i$, among the common RIs and RSs only between two adjacent $\gamma$-rate cells, the AV must pass through only one of them. 
%for all $\gamma$-rate cells on the bottom layer, 
We denote
the common RIs between two adjacent $\gamma$-rate cells, as well as the mid-points (as indicated in Fig. \ref{twolayer}(b)) of the common RSs that have no RIs on them, as \emph{core nodes}. The set of core nodes between $\gamma C_{j}^i$ and $\gamma C_{j+1}^i$ is denoted as 
$\mathcal{K}^{i,j}=\{k_1^{i,j},~k_2^{i,j},...,k_{I_{j}}^{i,j}\}$. By using the Dijkstra shortest-path algorithm, the intra-$\gamma$-rate-cell routing computes the minimum travel time $T_{n,m}^{i,j}$ from core node $k_n^{i,j-1}$ (between $\gamma C_{j-1}^i$ and $\gamma C_{j}^i$) to core node $k_m^{i,j}$ (between $\gamma C_{j}^i$ and $\gamma C_{j+1}^i$), for all $k_n^{i,j-1}$ in $\mathcal{K}^{i,j-1}$ and $k_m^{i,j}$ in $\mathcal{K}^{i,j}$ on $G_{\text{cell}}^j$\footnote{Unlike $G_{\text{BS}}$, computing the shortest paths is possible because the edge weights of $G_{\text{cell}}^j$ are defined to be the length of an RS divided by the effective speed from ESM.}.
In addition, let $T_{\text{opt}}^{i,j}$ denote the optimal travel time from the source location to core node $k_m^{i,j}$, through path $i$, and $T_0^{i,|P_i|}$ the optimal travel time from the source location to the destination location, through path $i$.

After inter-$\gamma$-rate-cell and intra-$\gamma$-rate-cell routing, dynamic programming is applied to compute $T_{\text{opt}}^{i,j}$, and eventually $T_0^{i,|P_i|}$.
Overall the two-layered routing scheme is summarized as follows:

%\begin{center}
%\begin{supertabular}{| p{.45\textwidth} |}
% \hline
%\textbf{Procedure 1: two-layered routing Scheme}~~~~~~~~~~~\\
% \hline
\textbf{Step 1}: Inter-$\gamma$-rate-cell routing. Find the set of all possible (top layer) paths on $G_{\text{BS}}$, $\mathcal{P}$, using BFS. 

\textbf{Step 2}: Intra-$\gamma$-rate-cell routing. Use the Dijkstra shortest-path algorithm to compute the minimum travel time $T_{n,m}^{i,j}, \forall~i,j,m,n$ and record the time with the corresponding path on the road network.

\textbf{Step 3}: Dynamic programming. $\forall\,i,\,j,\,m$, compute $T_{\text{opt}}^{i,j}=\min_n (T_n^{i,j-1}+T_{n,m}^{i,j})$. The shortest duration trip is then chosen to be $\min_iT_0^{i,|P_i|}$, with $P_{i^*}$ being the chosen top-layer path. The bottom-layer road paths can be obtained from the recorded results in Step 2.

\emph{Remark 1:} The necessary and sufficient condition for the two-layered routing scheme to have a solution is that the path set $\mathcal{P}$ obtained in the inter-$\gamma$-rate-cell routing in Step 1 is not empty. If no result is obtained at Step 1, one option is to reduce the rate threshold $\gamma$ until a path can be found.

\emph{Remark 2:} The proposed scheme is optimal, as shown next. 
Each route from the source location to the destination location corresponds to one distinct top-layer path ($P_i$), and in Step 1, all such top-layer paths, $\mathcal{P}$, are found. For $P_i$, in Step 3, the shortest travel time $T_0^{i,|P_i|}$ is obtained using dynamic programming. Comparing $T_0^{i,|P_i|}$ for all $i$, all results are exhaustively examined, and the route with minimum $T_0^{i,|P_i|}$ is the optimal solution to the problem.% in terms of trip duration.%The proof of optimality for this scheme can be referred to in \cite{zhang2018}.

% \emph{Remark 3:} Step 2 can be performed offline beforehand to reduce computation time, considering that preprocessing has already being applied in current vehicle navigations \cite{sanders2007}.
  % and try again to see if there are some other services available. 

\begin{figure}      \includegraphics[trim={3cm 17cm 8cm 20cm},clip, width=0.95\linewidth]{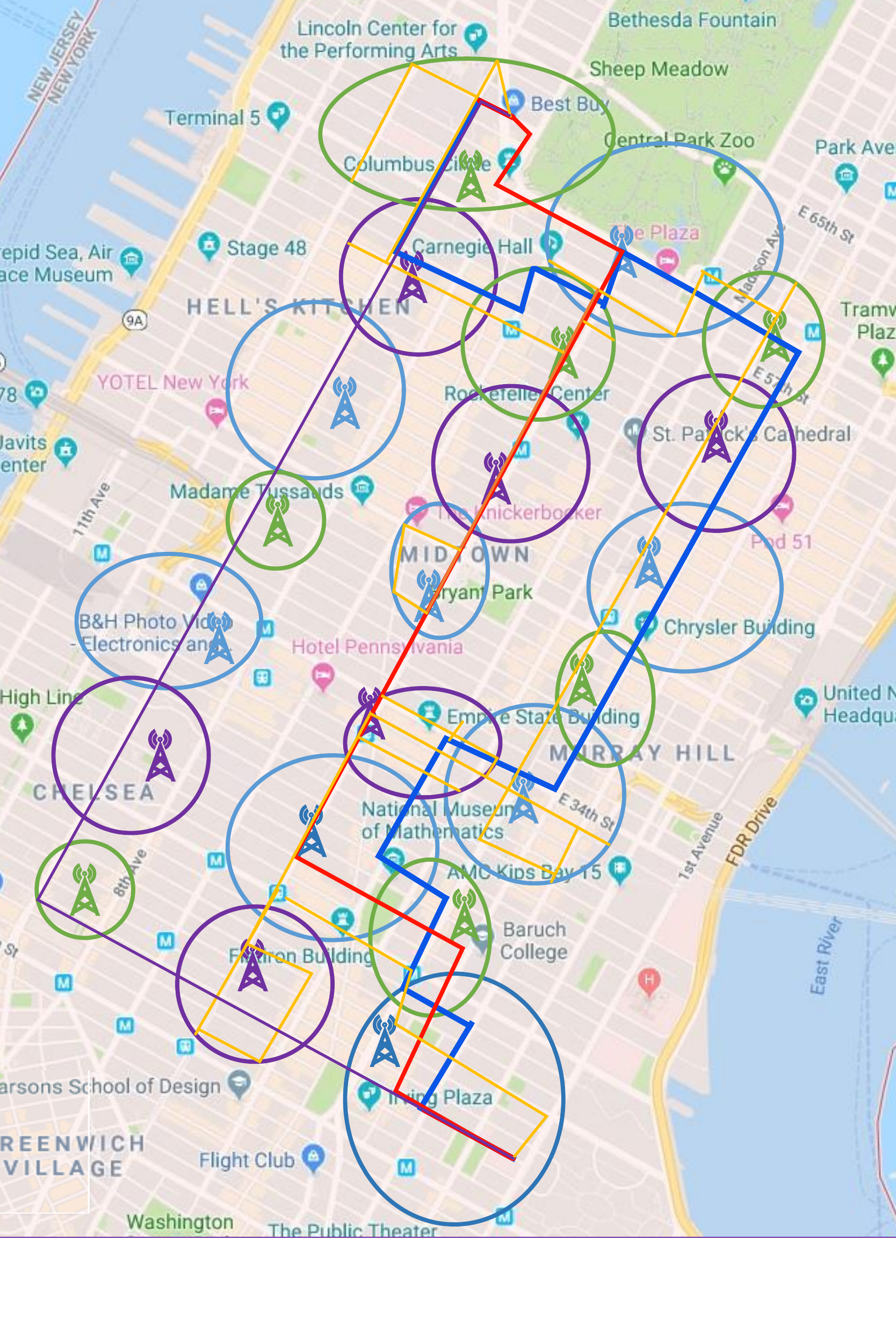}
        \vspace{-30pt}
                \caption{An example BS deployment in Manhattan and their $\gamma$-rate cells. The route in blue is computed by the ``Two Layer'' scheme; red by ``Greedy w/o CC''; yellow by ``Greedy s.t. CC''; purple by ``Shortest Time''. }
         
\label{route2}
\end{figure}

\subsection{Simulation Results}

We consider a Manhattan-like area as shown in Fig. \ref{route2}, where $A=11$ avenues and $S=51$ streets divide the area into a grid topology of identical rectangles\footnote{The use of a grid topology simplifies the greedy routing schemes used in the simulation.}. The corners of these rectangles are at the centers of the road intersections, the lengths and widths of which are $L=250$\,m and $W=100$\,m, respectively. $K=21$  BSs are uniformly deployed in this area and all are equipped with $N_t=128$ transmit antennas. Different from the realistic traffic patterns in Manhattan, we assume that all streets and avenues allow bidirectional traffic flows.

Given that there exists no other AV routing scheme subject to communication constraints, we compare the proposed two-layered AV routine scheme against two greedy routing approaches performed locally by the AV, one without communication constraints and one with communication constraints.

\subsubsection{Greedy routing without communication constraints}
In the scenario of a grid topology, typically, when the destination is located, say, northwest of the source, at each RI, an AV would choose the next RS that is (close to) either north bound or west bound, but neither east nor south bound, in order to travel the shortest distance. Specifically, at each RI, for all available RSs that lead closer to the destination, one is randomly chosen as the next RS. 

% Is there a better name for this algorithm?
\subsubsection{Greedy routing subject to communication constraints} In this scheme, while driving, the AV learns the availability of local RSs that could satisfy its communication constraints (from the information broadcast by BSs). Upon arriving at an RI, the AV will choose, among all the available local RSs that satisfy its communication constraints, the RS that leads closer to the destination.
In the worst case, the AV has to backtrack (i.e., make a U-turn) to a previous RI to choose a different RS (that satisfies the communication constraints).

Using specific values of ESM and $\gamma$, Fig. \ref{route2} shows the routes computed by different routing schemes. The `circle' around each BS represents its $\gamma$-rate cell. ``Greedy w/o CC" and ``Greedy s.t. CC" are the greedy methods introduced above, and ``Shortest Time" chooses the shortest-time path using the Dijkstra algorithm without communication constraints. For the route computed by ``Greedy s.t. CC", although it is mostly covered by $\gamma$-rate cells, its track shows backtracking behavior.

\begin{figure}       \includegraphics[ width=0.9\linewidth]{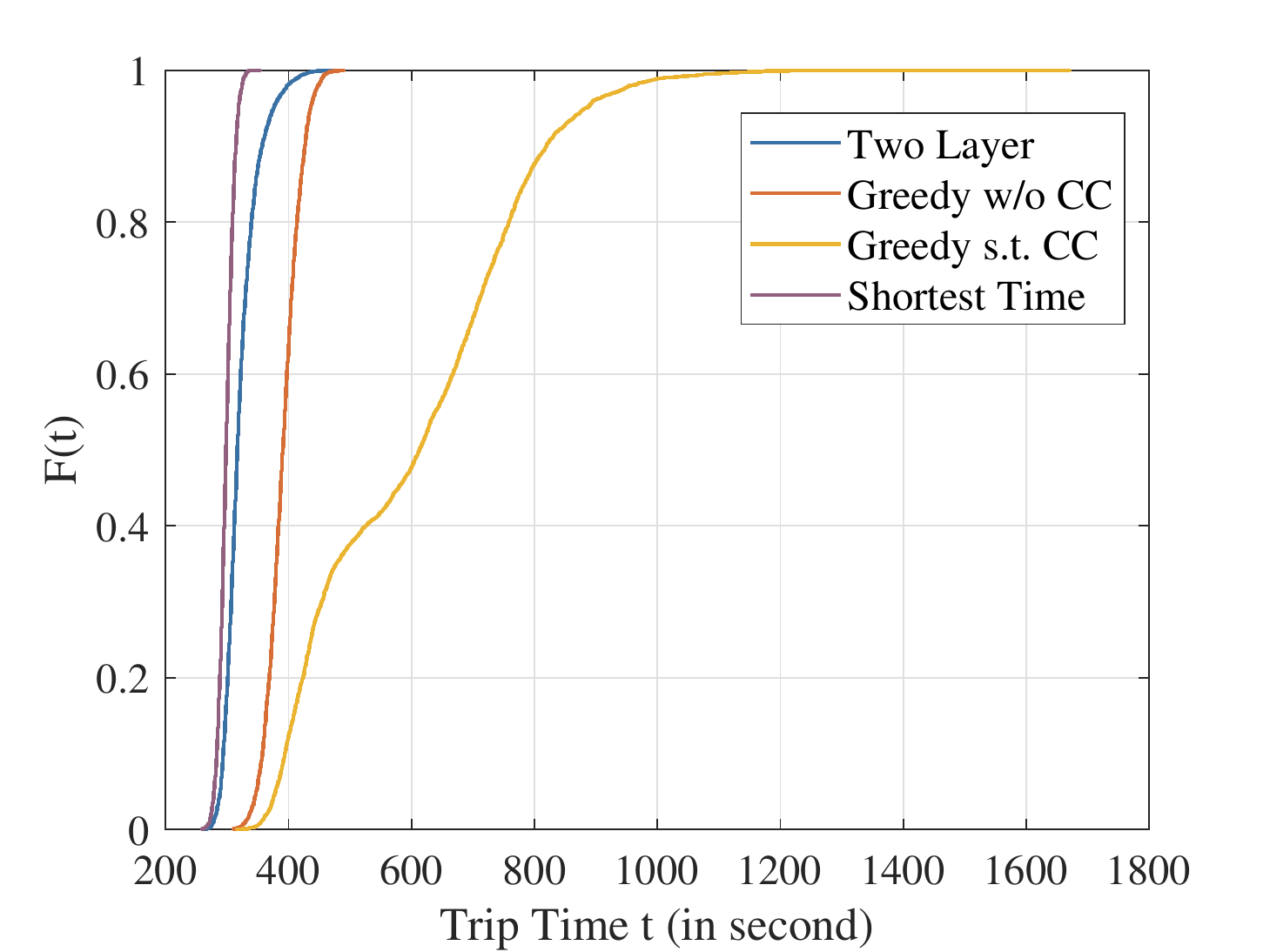}
     \caption{CDF of the trip duration ($\gamma$ is 55 Mbps).}
           
\label{cdf_time}
\end{figure}

%The variable speed limits for AVs in future Manhattan area are unknown, but o
% The proposed two-layered routing scheme is robust to the changes of ESM. To show this,
In the simulation, we randomly generate the effective speeds for all RSs 10,000 times, from the set \{10 m/s, 20 m/s, 30 m/s\}. These speeds are consistent with those used in \cite{3gpp2016}. The slot length in an LTE-A system, $\tau=1$ ms, is used as the channel measurement interval. Also, the Winner II path loss model for urban areas \cite{winnerII} is applied, and the transmission rates are computed using the method in \cite{papa2017}. Fig. \ref{cdf_time} depicts the CDF of the resulting trip duration for the different routing schemes with $\gamma$ = 55 Mbps. The results show that the proposed two-layered routing scheme can achieve a trip duration close to the shortest-time routing while satisfying the communication requirements. Note that, due to the potential backtracking in ``Greedy s.t. CC,'' an upper bound on the maximum number of RSs traveled is set. When this limit is reached, to reach the destination, the AV switches from ``Greedy s.t. CC'' to ``Greedy w/0 CC'', as noticed by the ``stepping'' appearance of the yellow curve in Fig. \ref{cdf_time}.
% (To avoid routing loops due to backtracking, a maximum number of RSs is set for  ``Greedy s.t. CC,'' passing which, greedy routing without communication constraint is performed, so its curve is less smooth.) 
\begin{figure} \includegraphics[ width=0.9\linewidth]{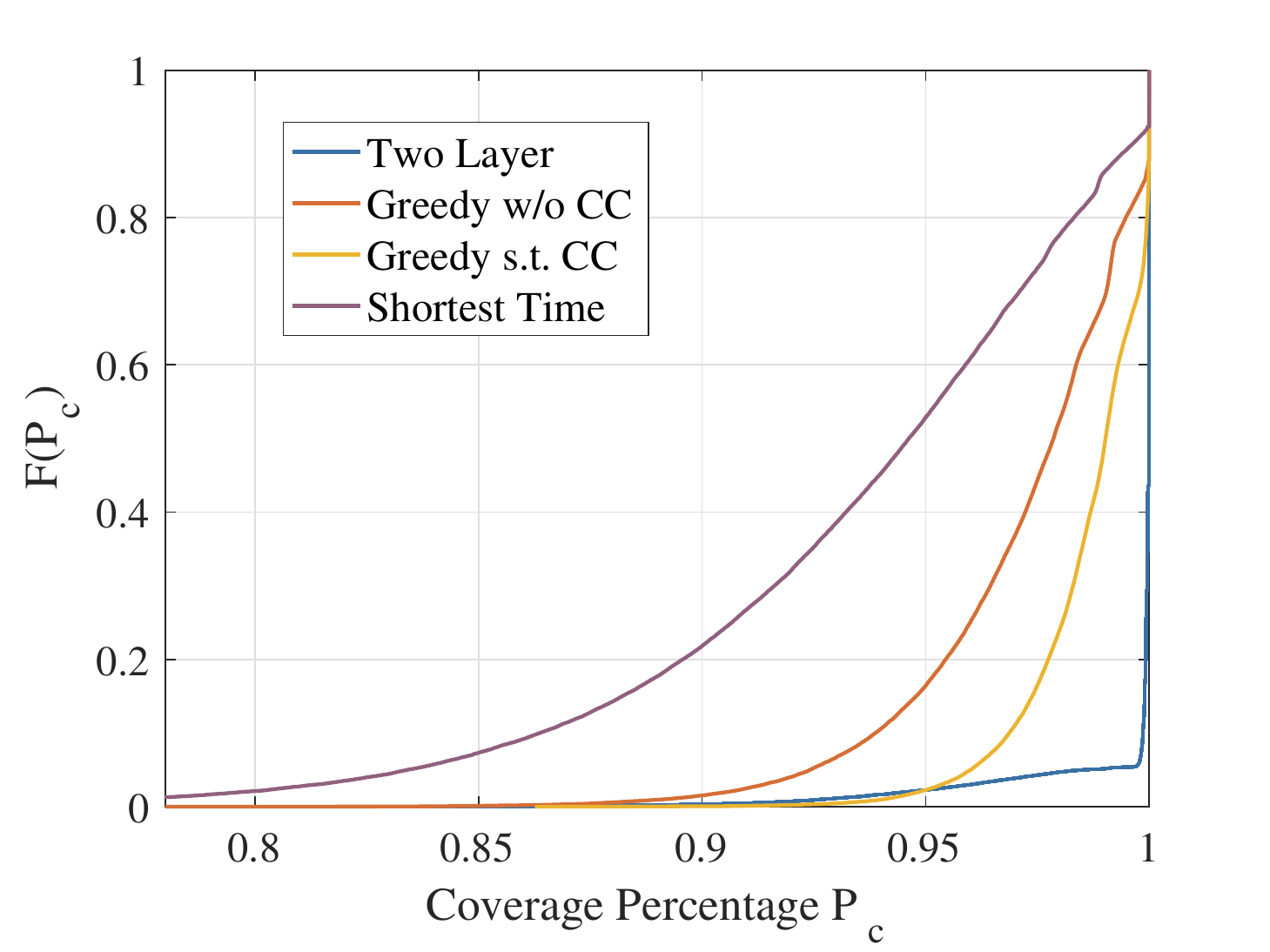}         \caption{CDF of the percentage of trip duration covered by the required communication ($\gamma$ is 55 Mbps).}
\label{cdf_coverage}
\end{figure}

Fig. \ref{cdf_coverage} depicts the CDF of the percentage of the trip duration that the AV's rate requirement is satisfied, $P_c$.
The results show that the proposed two-layered routing provides good communication coverage. In addition, because the ``Shortest Time" scheme usually chooses RSs with higher effective speeds, the Doppler effect affects the transmission rates, resulting in worse coverage than ``Greedy w/o CC.'' % (the only comparable algorithm in terms of coverage is ``Greedy w/o Commun.", but the trip duration for which can be long). 

\begin{figure}       \includegraphics[width=0.9\linewidth]{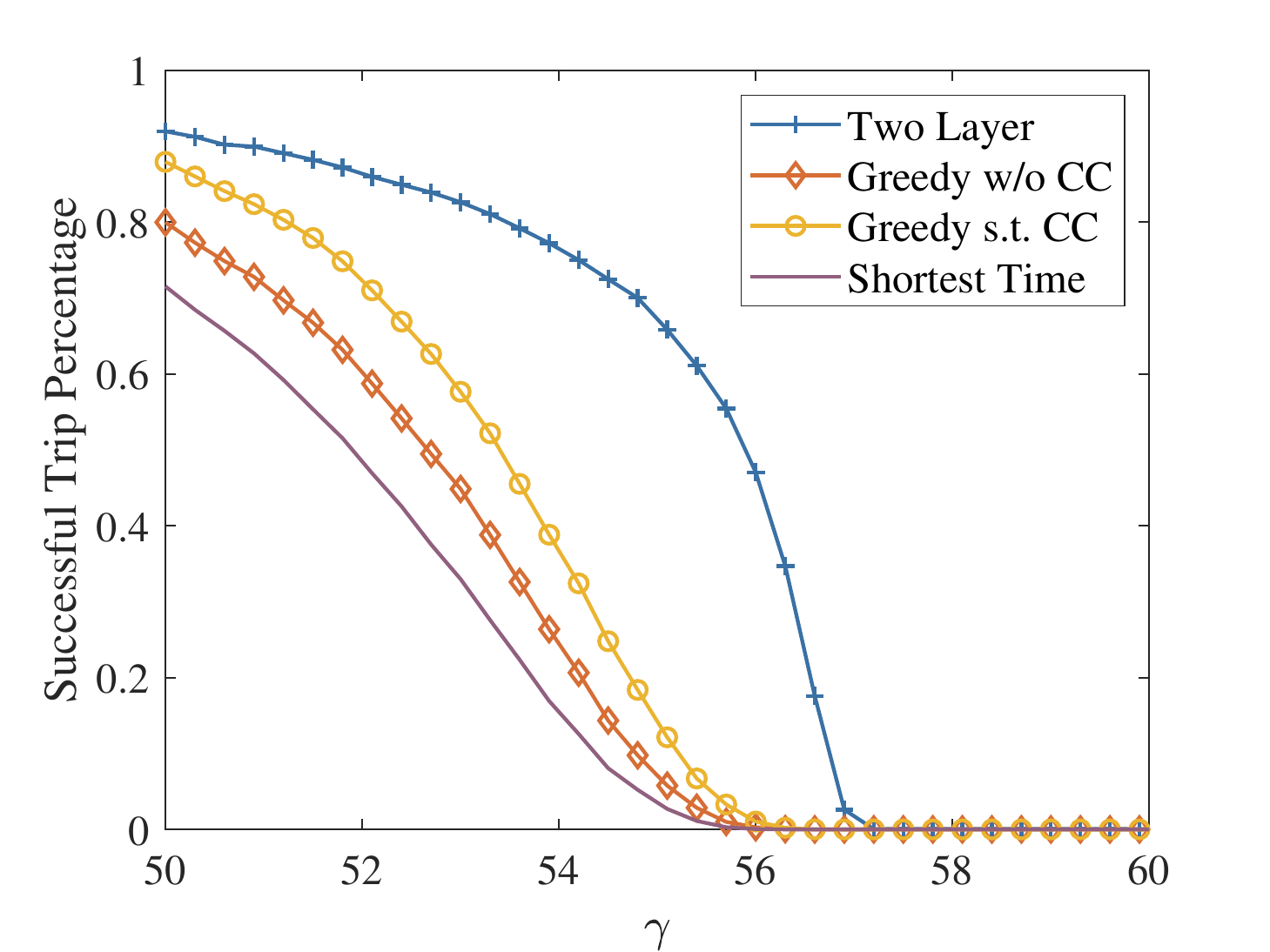}     \caption{Successful trip percentage w.r.t $\gamma$.}
\label{coverage_snr}
\end{figure}

\begin{figure}     \includegraphics[width=0.9\linewidth]{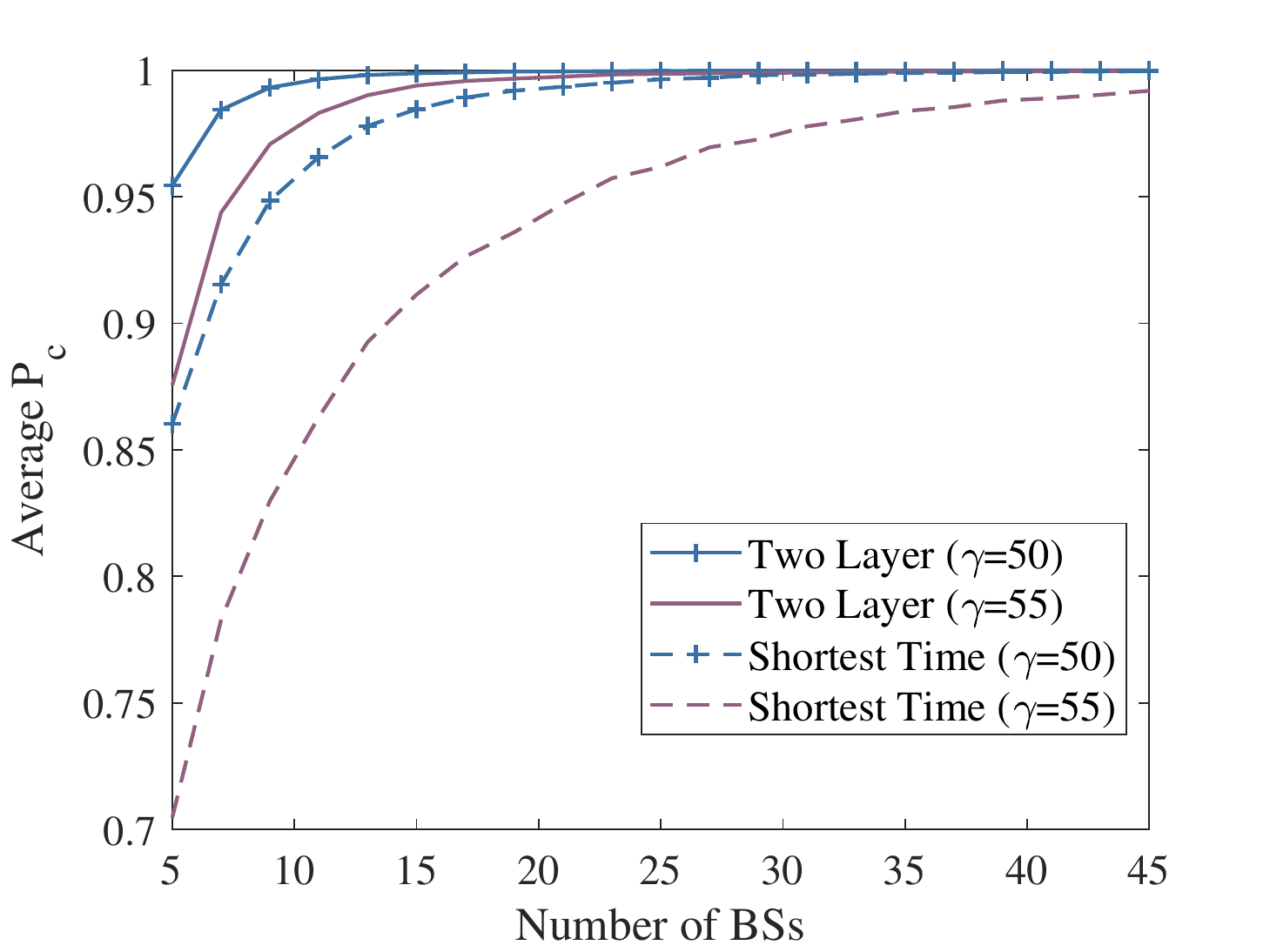}      \caption{Percentage of the trip covered by the required communication w.r.t the number of deployed BSs.}
\label{coverage_bs}
\end{figure}

\begin{figure}   \includegraphics[width=0.9\linewidth]{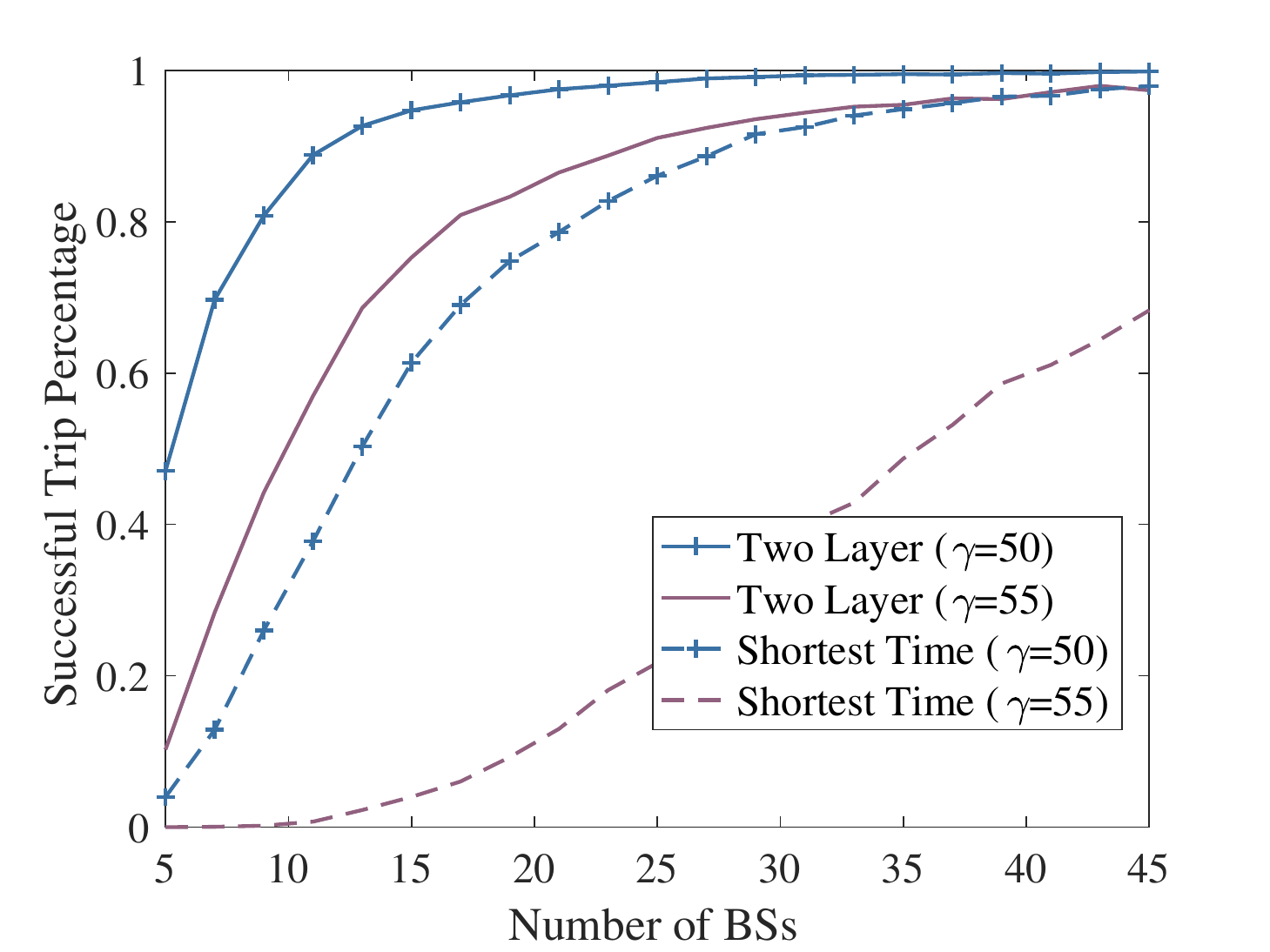}
     \caption{Successful trip percentage w.r.t the number of deployed BSs.}
\label{satisfactory_bs}
\end{figure}

Furthermore, we compute the metric``successful trip percentage'' defined as the probability that the target AV can find a route that satisfies the rate requirement along the entire trip. By varying the value of $\gamma$, the results, as shown in Fig. \ref{coverage_snr}, show that, with increasing $\gamma$, it becomes more difficult to find a satisfactory route; however, the two-layered routing scheme always outperforms the other methods using this performance metric. 
In addition, we vary the number of BSs in the Manhattan area. The average $P_c$ (average percentage of trip duration covered by the required transmission rates), as well as the successful trip percentage, are plotted in Figs. \ref{coverage_bs} and \ref{satisfactory_bs}, respectively. These two figures show that, to achieve a specified communication requirement, with CCR of AV, fewer BSs are required for deployment, but with a longer trip duration. 

% \begin{figure}         \includegraphics[trim={0cm 18cm 1cm 18cm}, clip, width=1\linewidth]{region.pdf}
%         %\vspace{-80pt}
%                 \caption{Potential  BS deployment in Manhattan and their 28 Mpbs rate region}
% \label{orignial}
% \end{figure}

\section{Communication-constrained Traffic Control (CCTC)}\label{control}

%\subsection{Joint AV Traffic Control and Communication Planning}
In the last section, we considered the routing of a single AV and ignored the impacts of other AVs, even though a BS can serve multiple AVs simultaneously. From the viewpoint of traffic control, maximizing the throughput capability of a road network is one primary goal of adopting AVs \cite{boyles2018}. However, to achieve the optimal CCTC of maximizing the road-network throughput subject to communication constraints, several interrelated issues remain to be studied.
%, which include but are not limited to: 1) invent methods to jointly route AVs to prevent that certain BSs admit more AVs than it can support; 2) invent admission control strategy for BSs to reduce the burden of admission control for adjacent cells and AV admission delay; 3) invent methods to evacuate AVs in a short time when congestion or other accidents occur; 4) investigate and improve the road-network throughput upper bound by studying the fundamental relationship between the cellular systems performance metrics and AV movement control, and optimize the AV operation parameters to approach this upper bound; 5) coordinate BSs, including their  communication resources, to optimize the traffic flows across multiple cells.     

In this paper, we study two specific problems: (1) in one-cell scenarios, given the constrained communication resources, we investigate the number of AVs that can be served simultaneously by the BS as a new KPI for traffic control, and use the result to derive the optimal AV speed for maximizing the ``sum traffic flow'' into/out of the cell, and (2) in multi-cell scenarios, we study a traffic flow capacity mismatch problem across different cells along the same road, and propose a spectrum re-balancing solution to address the issue.  Numerical results demonstrate that the one-cell and multi-cell solutions can be applied in sequence to maximize the road-network throughput.  

\begin{figure}     \includegraphics[width=1\linewidth]{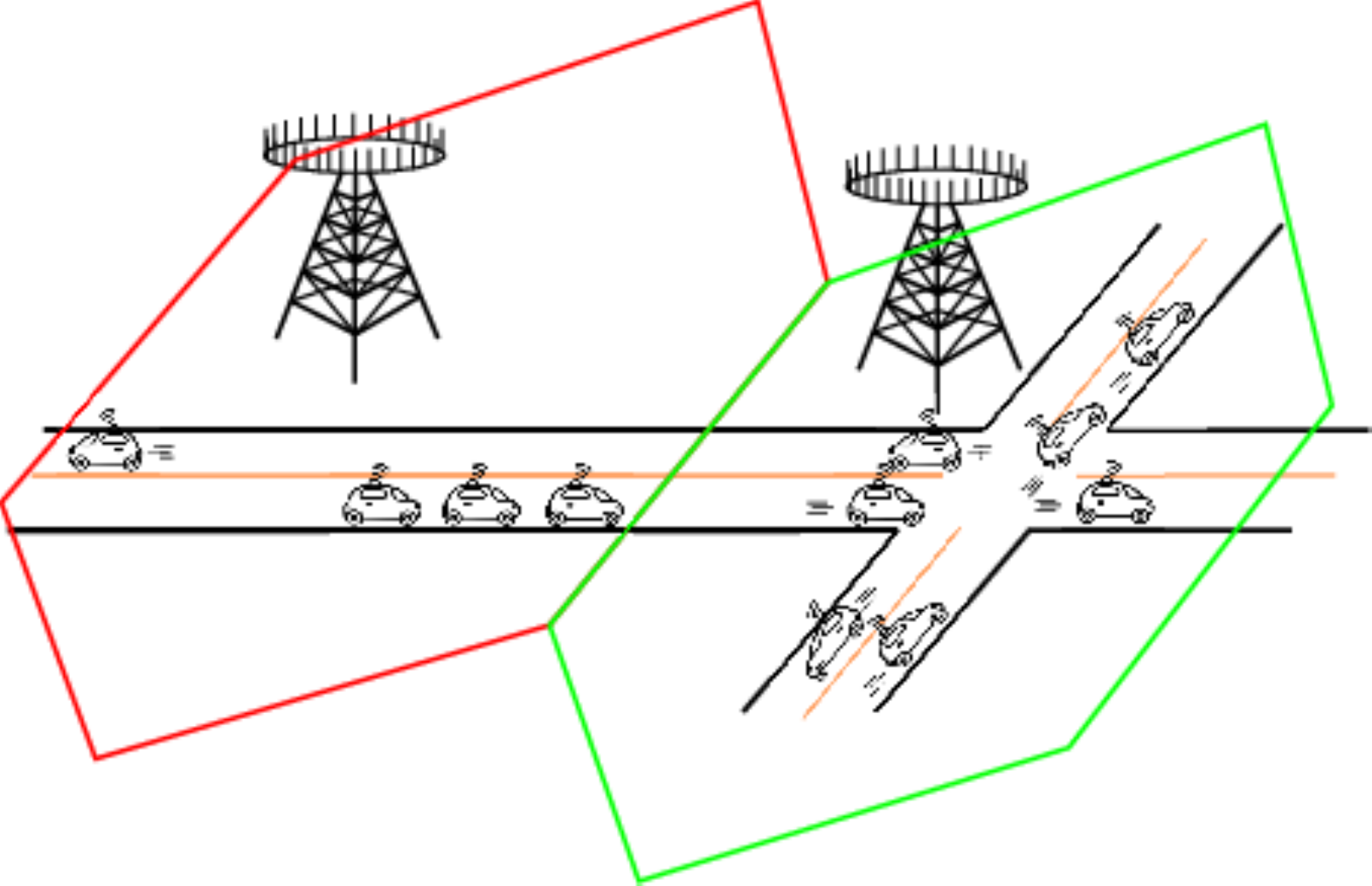}
 \caption{Scenario of traffic flow capacity mismatch across cells}
\label{traffic}
\end{figure}
\subsection{Speed Optimization within Single Cell} \label{sec:onecell}

% In this subsection, we first introduce a cellular system KPI for the purpose of measuring the capability of serving AVs for traffic control. The cell sum traffic flow can be derived using this KPI. Then, assuming every AV drives at the same speed, we prove the existence of an optimal AV effective speed for each cell to maximize the cell's sum traffic flow. 

As discussed in \cite{3gpp2018_2}, 
to enable high LoA for AVs, certain minimum communication requirements of reliability, transmission rate, and latency, denoted by $\Gamma$ = [$\mathcal{R}_0, \mathcal{G}_0, \mathcal{A}_0$], must be satisfied for each AV\footnote{The higher the LoA, the more stringent the communication requirements $\Gamma$.}.
%such that, the communication reliability $\mathcal{R}$ should be larger than $\mathcal{R}_0$, the transmission rate $\mathcal{A}$ should be higher than $\mathcal{A}_0$, and the latency $\mathcal{L}$ should be smaller than $\mathcal{L}_0$. 
We propose the following KPI, connecting $\Gamma$ with AV traffic control, to indicate the traffic control performance of a single cell. 
\begin{quote}
 %Note that the area outside the boundaries of real physical roads is not counted.
$\mathcal{N}_{\Gamma}^v(\mathcal{B})$, which denotes the largest number of AVs that a BS can control simultaneously subject to the communication requirements of $\Gamma$, where AVs are uniformly distributed within the BS's coverage area, all AVs move at speed $v$, and $\mathcal{B}$ is the number of available frequency channels in this cell.
\end{quote}
Note that $\mathcal{N}_{\Gamma}^v(\mathcal{B})$ monotonically increases with $\mathcal{B}$ and decreases with $v$, given the same $\Gamma$.
The decrease with speed results from the fact that more communication resources would need to be consumed for each AV traveling at higher speeds to satisfy $\Gamma$. Given the the same amount of available resources, fewer AVs could be controlled when traveling at higher speeds. For simplicity, we do not consider optimal channel allocation across channels; then, for identically distributed frequency channels within a cell, there exists a value $\mathcal{B}_m$ such that the following linearity property holds.
\begin{quote}
$\forall~\mathcal{B}_1+\mathcal{B}_2<\mathcal{B}_m,~ \mathcal{N}_{\Gamma}^v(\mathcal{B}_1+\mathcal{B}_2)=\mathcal{N}_{\Gamma}^v(\mathcal{B}_1)+\mathcal{N}_{\Gamma}^v(\mathcal{B}_2)$
$\forall~\mathcal{B}<\mathcal{B}_m, ~\mathcal{N}_{\Gamma}^v(\mathcal{B})=\mathcal{B}\mathcal{N}_{\Gamma}^v(1)$
\end{quote}
When there is only one frequency channel, a BS can still control multiple AVs by using
time-division multiplexing (TDM) or frequency-division multiplexing (FDM) to divide the channel into multiple subchannels, as well as by 
using MIMO. Here, we assume this linearity property always holds. %$\mathcal{N}_{\Gamma}^v(\mathcal{B})$ is a non-negative integer.  

% , for steady traffic flows. 
%In order to derive the formula for $\mathcal{F}_m$,  and 

For simplicity, we assume that, within a cell, there is no lane merging and no dead ends, so that the numbers of lanes, $\mathcal{L}$, into and out of a cell are identical. (For instance, in Fig. \ref{traffic}, for the red cell, $\mathcal{L}=2$; and ,for the green cell, $\mathcal{L}=4$.). We define BS Coverage $\mathcal{C}$ as the total surface area of the roads that the BS covers. The width of each lane is assumed to be $\mathcal{W}$; then, the total length of all lanes in this cell is $\dfrac{\mathcal{C}}{\mathcal{W}}$. 

Let $\mathcal{D}(v)$ denote the minimum distance between two AVs within the cell set by the transportation authority subject to effective speed $v$. For a steady flow of traffic, we consider that AVs are uniformly distributed in the BS's coverage area with speed $v$. Then, in theory, the distance between two consecutive AVs would be $\max\{\mathcal{D}(v), \dfrac{\mathcal{C}}{\mathcal{W}\mathcal{N}_{\Gamma}^v(\mathcal{B})}\}$. The result of dividing this distance by $v$ becomes the average inter-arrival time between two consecutive AVs entering (or leaving) the cell. We define {\em cell sum traffic flow}, $\mathcal{F}_m$, to be the maximum number of AVs that is allowed to enter or leave a cell per minute. Then, by applying the equation on p. 319 in \cite{friedrich2016}, \begin{align}
    \mathcal{F}_m=\dfrac{\mathcal{L}v}{\max\{\mathcal{D}(v), \dfrac{\mathcal{C}}{\mathcal{W}\mathcal{N}_{\Gamma}^v(\mathcal{B})}\}}.
    \label{eq:Fm}
\end{align} 
%If the cell is small and control only part of the roadside unit, then the
% Considering that AV platooning is being employed so that AV distances can be really small, we can set $\mathcal{D}(v)=0$. 
In practice, $\mathcal{D}(v)$ is small and can be eliminated from Eq. (\ref{eq:Fm}).
Therefore, the average distance between consecutive AVs becomes $\dfrac{\mathcal{C}}{\mathcal{W}\mathcal{N}_{\Gamma}^v(\mathcal{B})}$. The result of dividing this average distance by $v$ is the average inter-arrival time between two consecutive AVs entering or leaving the cell. Then, \begin{align}\label{eq:flow}
    \mathcal{F}_m=\dfrac{\mathcal{L}\mathcal{W}}{\mathcal{C}}\mathcal{N}_{\Gamma}^v(\mathcal{B})v.\end{align}
Note that (\ref{eq:flow}) holds even for non-steady-state flows, where the inter-arrival time between AVs entering a cell is not uniformly distributed.
%By default, there is always a sufficient number of AVs queuing up to enter these cells. %Also, the road conditions are identical for different RSs in one cell, such that their corresponding effective speeds are identical. 
Then, for the one-cell case considered here, the following lemma shows that each cell has an optimal speed for maximizing the road-network throughput in that cell.

\emph{Lemma~1}: 
% When there is one AV controlled by a BS, let the AV's maximum speed without violating the communication requirements $\Gamma$ be $v_m$, where $0<v_m\le v_l$. 
% Besides, there are sufficient AVs waiting to enter this cell. 
Given the fixed amount of communication resources in a cell and the minimum communication requirements $\Gamma$ for a specified LoA, there exists an optimal speed, and hence an optimal number of AVs, such that the road-network throughput in this cell is maximized when all AVs travel at this speed.
\begin{proof} Given a fixed amount of communication resources in a cell, the relationship between $\mathcal{N}_{\Gamma}^v(\mathcal{B})$ and the AV speed $v$ is a unique function, and $\mathcal{N}_{\Gamma}^v(\mathcal{B})$ monotonically decreases with $v$.   $\mathcal{N}_{\Gamma}^v(\mathcal{B})$ peaks at $v=0$.  
Also, there exists a speed threshold $v_m$ such that, when all the AVs in the cell travel at a speed above this threshold, they become uncontrollable, i.e., $\mathcal{N}_{\Gamma}^v(\mathcal{B})$ = 0, when $v>v_m$.
% Also, according to the definition of  $v_m$, when $v>v_m$, $\mathcal{N}_{\Gamma}^v(\mathcal{B})=0$.
Therefore, $\mathcal{N}_{\Gamma}^v(\mathcal{B})v$ is upper bounded by $\mathcal{N}_{\Gamma}^0(\mathcal{B})v_m$.
Because $\mathcal{L},\,\mathcal{W},$ and $\mathcal{C}$ are constants, there is an AV speed $v^*\le v_m$ that maximizes $\mathcal{F}_m=\dfrac{\mathcal{L}\mathcal{W}}{\mathcal{C}}\mathcal{N}_{\Gamma}^v(\mathcal{B})v$ ($\mathcal{F}_m^*$), and $\mathcal{N}_{\Gamma}^{v^*}(\mathcal{B})$ is the corresponding optimal number of AVs. For this cell, this value of sum traffic flow $\mathcal{F}_m^*$ is its maximum road-network throughput.
\end{proof}

Lemma 1 only proves the existence of an optimal AV speed.
To derive the optimal AV speed, it becomes necessary to obtain the unique function between $\mathcal{N}_{\Gamma}^v(\mathcal{B})$ and AV speed $v$. This function depends on the technology employed by the BS, which can be obtained by using either statistical methods or model-based methods. In Subsection \ref{appdemon}, a model-based method is introduced for a Time Division Duplex (TDD) BS. Also note that the optimal AV speed can be used for constructing the ESM for the CCR problem from the previous section, and this speed could differ in different cells.

\subsection{Spectrum Balancing across Multiple Cells}

For reasons such as severe channel fading and/or a high number of AVs, it is possible that the available communication resources within a cell are being exhausted so that it will not be able to support the desired LoA when more AVs enter the cell. Fig. \ref{traffic} depicts such a scenario where the green cell does not have enough communication resources to accommodate more AVs coming from the red cell, so AVs may need to back up at the cell boundary.

For a road crossing multiple cells, (communication constrained) traffic flow capacity mismatch between adjacent cells can lead to traffic congestion in a cell that supports less traffic flow. Among all the cells that a road travels across, we term the cell with the least amount of traffic flow capacity {\em bottleneck cell}. The overall traffic flow of a road is then constrained by the capacity of the bottleneck cell. Therefore, to avoid congestion, ideally, all the cells along a road should support the same traffic flow capacity. We provide Lemma 2, which shows that dynamic channel allocation across different cells provides a solution to mitigate the negative effects of the bottleneck cells. For instance, in current cellular systems, adjacent cells often use different frequencies to avoid inter-cell interference. Therefore, given the total number of channels $\mathcal{B}_0$ over an area, more channels should be assigned to the bottleneck cells to increase their allowed traffic flows, and hence the overall throughput of the road crossing multiple cells.

Consider the following scenario for Lemma 2. An area containing $L_r$ one-way roads (a two-way road can be considered as two one-way roads) is fully covered by $N_e$ hexagonal cells that do not overlap. Within this area, in total, there are $\mathcal{B}_0$ frequency channels, and within a cell, each channel can control the same number of AVs. %, and each AV is controlled via at least one frequency channel. 
No two cells in this area share the same channel. BS $i$ assigns $\mathcal{B}_{ij}$ frequency channels for the AVs on road $j$, such that the maximum traffic flow on this section of road $j$ within cell $i$ is $\mathcal{F}_{ij}$. $\sum_{i,j}\mathcal{B}_{ij}=\mathcal{B}_0$. 
We consider FDM
and assume that each frequency channel can be divided into a large number of subchannels to control multiple AVs. Also, it is allowed to assign a fraction of the channels to a cell.
%For the sake of stating the proof of Lemma 2, we allow $\mathcal{B}_{ij}$ to be a fractional number, where the fractional part $\mathcal{B}_{ij}-\lfloor\mathcal{B}_{ij}\rfloor$ represents a certain number of subchannels. This is valid when it is  
%where $\mathcal{B}_{ij}-\lfloor\mathcal{B}_{ij}\rfloor$ should always be an integer times the bandwidth fraction of one subchannel in one frequency channel, representing this integer number of subchannels. The validity of the proof does not depend on this fractional nature.
The proof for TDM is similar. %DOES THE PROOF DEPEND ON THE FRACTIONAL NATURE?  IF IT DOES YOU NEED TO SAY SOMETHING.

%Sup divisible to smaller frequency channels by any integers: $\mathcal{N}_{\Gamma}^v(1)=M\mathcal{N}_{\Gamma}^v(\dfrac{1}{M})$ where $M$ is a positive integer.

\emph{Lemma~2}: Given $\mathcal{B}_0$ frequency channels to control AVs over an area containing $L_r$ one-way roads, a necessary and sufficient condition for deciding assignments $\mathcal{B}_{ij}$  to achieve Pareto-optimal road throughput among the $L_r$ roads in this area is that, % should be assigned such that, 
%for each road $j$, the allowed maximum traffic flows in different segments of the road controlled by different BSs should be equal. That is,
for each road $j$, $\mathcal{F}_{i_1j}=\mathcal{F}_{i_2j}$ for all BSs $i_1$ and $i_2$ whose cells road $j$ travels across.
\begin{proof}
Randomly assign all $\mathcal{B}_0$ channels across different roads and cells. 
Suppose road $j$ crosses a sequence of cells $\mathcal{S}_j=~<s_1, s_2\cdot\cdot\cdot s_{I(j)}>$ in turn.
%The number of frequency channels in cell $i$ for this road is $\mathcal{B}_{ij}$, and then the allowed traffic flow for this RS can be determined to be $\mathcal{F}_{ij}$. 
Then, the traffic flow capability on this road should be $\mathcal{F}_j=\min_{s_i\, \text{in}\,{\mathcal{S}_j}} \mathcal{F}_{ij}$, which is constrained by the cell with the smallest $\mathcal{F}_{ij}$. 
After the initial assignment, repeatedly shift channels or their subchannels from the cell with highest $\mathcal{F}_{ij}$ to the cell with lowest $\mathcal{F}_{ij}$, one subchannel at a time, until $\mathcal{F}_{i_1j}=\mathcal{F}_{i_2j}$ (if $\mathcal{F}_{i_1j}=\mathcal{F}_{i_2j}$ is not achievable due to the fact that the number of AVs has to be an integer, then at least $\mathcal{F}_{i_1j}\approx\mathcal{F}_{i_2j}),$ $~\forall~s_{i_1},\, s_{i_2}$ in $\mathcal{S}_{j}$. Repeat this process for all $L_r$ roads in this area.
%there should be a set of contiguous sections of road on road $j$, $\mathcal{T}_j$, whose indexes are denoted as $i_s,i_s+1,\cdot\cdot\cdot,i_e$, such that $\forall\,i_s\leq i\leq i_e,\,\mathcal{F}_{ij}=\mathcal{F}_{i_sj}$, and $\mathcal{F}_{ij}\leq \mathcal{F}_{kj}$ for $k\not\in \mathcal{T}_j$. Then, $\mathcal{F}_j=\mathcal{F}_{i_sj}$. $\mathcal{F}_j$ can be improved simply by reducing $\mathcal{B}_{(i_s-1)j}$, or $\mathcal{B}_{(i_e+1)j}$, or both, by certain amounts, and split these amounts to all sections of road in $\mathcal{T}_j$ such that $\mathcal{F}_{(i_s-1)j}=\mathcal{F}_{(i_s)j}=\cdot\cdot\cdot=\mathcal{F}_{(i_e)j}=\mathcal{F}_{(i_e+1)j}$. Repeat this to improve $\mathcal{F}_j$  until $\mathcal{F}_{i_1j}=\mathcal{F}_{i_2j}\, \forall~s_{i_1} s_{i_2}\in \mathcal{S}_j$. Then, repeat the above process for all other roads in this area, 
%until for any road $j'$, $\mathcal{F}_{i_1j'}=\mathcal{F}_{i_2j'} ~\forall~s_{i_1},\, s_{i_2}\in \mathcal{S}_{j'}$. 
With these adjustments, the road throughputs of $L_r$ roads all become higher.

 To further improve the throughput for road $j$, channels or subchannels need to be reassigned from the other roads. Because each channel can control the same number of AVs within a cell, if channels are switched between two roads, the number of channels for each of the roads stays the same, so that their allowed traffic flows stay the same. However, borrowing channels or subchannels is always at the cost of the other roads' throughputs. Thus, as long as $\mathcal{F}_{i_1j}=\mathcal{F}_{i_2j}, \,\forall~s_{i_1},\, s_{i_2}\in \mathcal{S}_j$, and $\forall\,j$, it is impossible to improve the throughput of a certain road without compromising another, and hence the throughputs for the $L_r$ roads are Pareto-optimal.
\end{proof}

% Note that Lemma 2 applies to the cases where the roads are not straight. 
Note that when frequency reuse is allowed across cells such that two cells not adjacent to each other may share the same spectrum, or when different subchannels can control different numbers of AVs, ``$\mathcal{F}_{i_1j}=\mathcal{F}_{i_2j}$ for all BSs $i_1$ and $i_2$" in Lemma 2 is only a necessary condition for achieving Pareto-optimal road throughput. %Also, for FDM, one channel could control multiple 

\emph{Remark 3:} From the two lemmas, it is possible that two adjacent cells may have different optimal speeds and different optimal distances between AVs. In practice, the movement control services running in the infrastructure would foresee such situations to gradually slow down or speed up AVs to avoid any abrupt change of speed when crossing cell boundaries along the road. 
% an AV would have to follow the ``rule" in that cell, and experience a sudden change of speed, distance between AVs, and the number of frequency channels for its control.  
% This setting is different from the current road network where each vehicle makes its own decision subject to a universally-applied traffic rule. %This sacrifice of driving flexibility is a cost for achieving maximal road-network throughput in traffic control with communication constraints.  

\subsection{Illustrative Scenarios}\label{appdemon}

\subsubsection{Derivation in one-cell scenarios} We consider the speed optimization problem in a single-cell slotted TDD MIMO scenario, where the duration of each slot is $T_{\text{slot}}$. By periodically sending $T_{\text{pilot}}$-length (in sec) pilots to measure the channels, the BS can obtain \emph{accurate channel information}; this information is critical for satisfying the downlink communication requirements $\Gamma$. In the downlink, the BS sends periodic messages to each AV at a repetition rate of $\lambda_m$, which means $\lambda_m$ messages are sent per second. 
%(I DONT UNDERSTAND THIS SENTENCE - THE ONE RIGHT BEFORE THIS)
The duration of one message is $T_{m}$, in sec, which is an integer multiple of $T_{\text{slot}}$. Because of the linearity property discussed above in Subsection \ref{sec:onecell}, we focus on one frequency channel in the analysis, and use TDM for sharing the channel.

Given accurate channel information, and using one frequency channel to send the $T_{m}$-length messages, $L$-user-MIMO technology is applied in the BS, such that, at every downlink time slot, $\Gamma$ can be satisfied for a maximum of $L$ AVs. Note that, in order for the repetitive message transmissions to be successful, the message inter-arrival time, $1/\lambda_m$, should be larger than the duration of one message, $T_{m}$; then, $\lambda_mT_{m} < 1$.  %In addition, the traffic flow in this cell is steady, such that all AVs travel at an identical speed, and have equal distance between adjacent AVs. 

%A reference AV speed $v_0$ is given to the cell such that it corresponds to a minimum channel measurement frequency $f_0$ to obtain accurate channel information and to meet the communication requirements $\Gamma$. $f_0$ should increase with $v_0$ but the exact relationship depends on the spatial channel coherence. 

To obtain channel information that is sufficient and accurate enough to satisfy the $\Gamma$ requirements, each AV must periodically send pilots to measure its channel at a maximum time interval of $T_v\approx \dfrac{1}{\alpha f_D(v)}$, where $f_D(v)$ is the AV's Doppler frequency and $\alpha$ is a scaling factor larger than 1. $f_D(v) = \dfrac{v}{c}f_c$, where $f_c$ is the carrier frequency and $c$ is the speed of light. The `$\approx$' is because in practice $T_v$ should be an integer multiple of $T_{\text{slot}}$. 
In the frequency domain, one channel contains multiple subchannels, and during the $T_{\text{pilot}}$-length measurement of one AV, this AV's pilots are sent over all these subchannels; % to obtain their channel information for the $L$-user MIMO beamforming at the BS; 
however, since these pilots are sent at the same time during $T_{\text{pilot}}$, we focus only on the time domain in the analysis below. Note that, within one frequency channel, TDM is used for multiplexing; dividing the frequency channel into multiple subchannels is for OFDM rather than for FDM. %(I STILL THINK THIS IS CONFUSING BECAUSE YOU HAVE SUBCHANNELS AND TDM AND MIX THEM UP) %When the channel estimation error is caused mostly by channel aging, such as the model in \cite{papa2017}, the transmission rate is usually a function of the correlation between the measured channel and the real channel $\rho_0$, which is a function of Doppler frequency $f_D(v)$. % For instance, in a Rayleigh fading scenario, $\rho_0=J_0(2\pi\dfrac{f_D}{f_0})$, where $J_0(\cdot)$ is the zeroth-order Bessel function.
%Also, in wireless scenarios, communication reliability is defined to be  solely dependent on the outage probability for one-hope transmission (see Subsection II.C in \cite{park2019} and Subsection III.B in \cite{ge2019}), which is usually also a function of $\rho_0$ \cite{ firag2011, coskun2012}. In addition,  latency depends on, among some other factors, the probability of retransmissions, thus also the outage probability. Therefore, when retransmission caused by outage is the limiting factor for latency, or when the latency requirement is always satisfied, roughly, to guarantee the same $\rho_0$ and thus $\Gamma$ for speed $v$, channel measurement should be performed at a minimum frequency of $\alpha f_D(v)$. %$f_v$. %Let the minimum slot interval be $\tau_n$, then in unit time the number of channel measurement pilots should be smaller than the number of such minimum slot and $f_v\leq\dfrac{1}{\tau_n}$ and $v\leq \dfrac{v_0}{f_0\tau_n}$, so $v_l$ should be set smaller than  $\dfrac{v_0}{f_0\tau_n}$.

\begin{figure}     \includegraphics[width=0.8\linewidth]{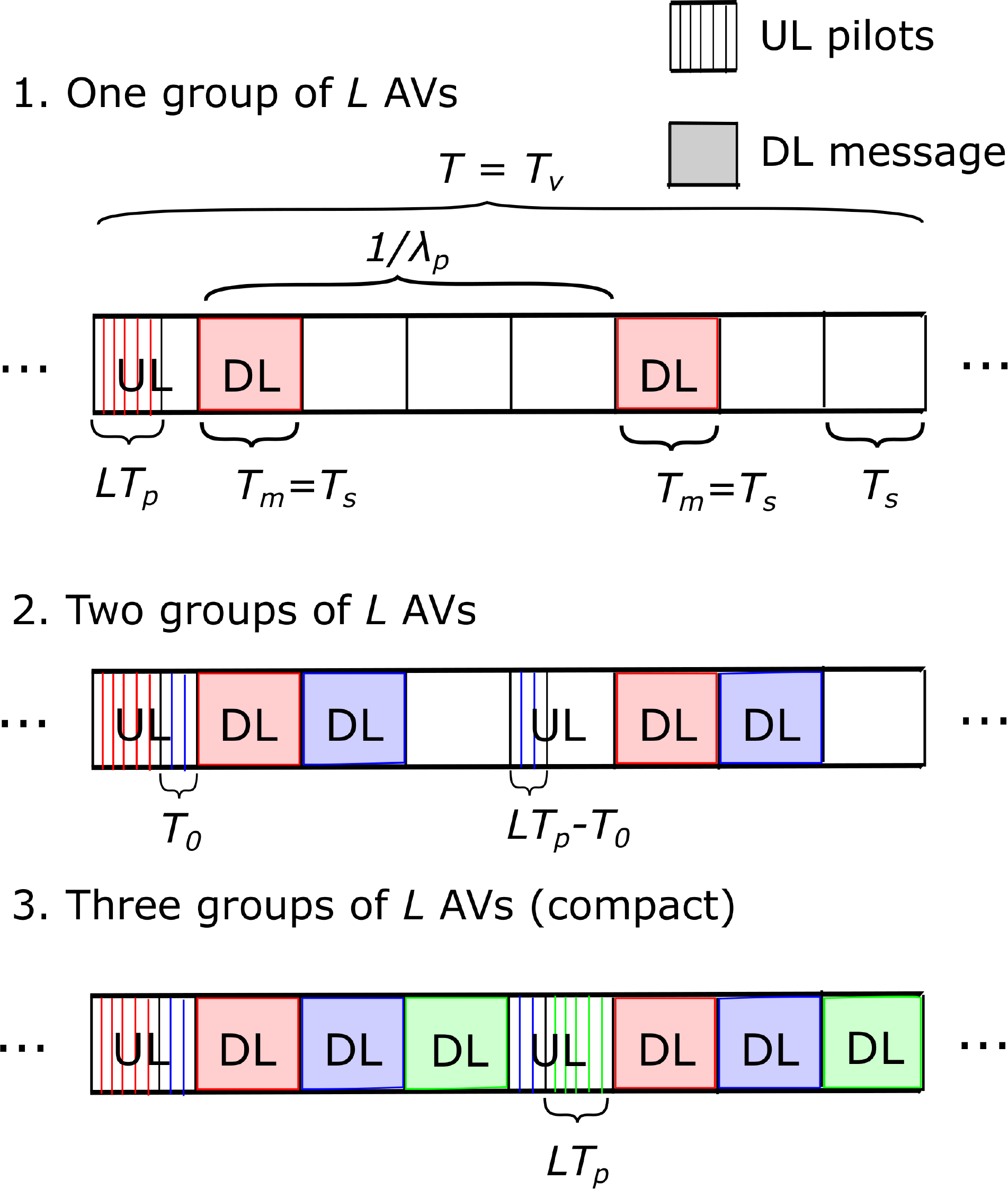}
 \caption{TDD MIMO downlink (DL) and uplink (UL) time slots filling-up during $T$. For the same color, there are two types of rectangular symbols with different patterns, respectively representing $T_{m}$-length downlink messages and uplink pilots for one group of $L$ AVs. %Note that they have different patterns for their symbols in the figure, the former's pattern tilting down to the left and the latter' tilting to the right. 
 Red, blue, and green represent the first, second, and third group of $L$ AVs.}
\label{Fig: slots}
\end{figure}

From the discussions in the two paragraphs above, in the time domain, the slots are occupied by the AVs, sending periodic channel measurement pilots in the uplink, in turn, as well as receiving messages from BSs periodically in the downlink, with at most $L$ AVs sharing one downlink slot. Furthermore, within a time period $T$ ($T\gg T_{m}$), each AV would require roughly $T/T_v$ pilots, such that $N_{\gamma}^v(1)$ AVs would require $N_{\gamma}^v(1)(T/T_v)$ pilots, which take $\lceil\dfrac{N_{\gamma}^v(1)(T/T_v)T_{\text{pilot}}}{T_{\text{slot}}}\rceil$ slots. Also, for downlink message transmission using $L$-user MIMO, within a time period $T$, one group of $L$ users would require $\lambda_mT$ $T_{m}$-length slots for sending messages, such that $N_{\gamma}^v(1)$ AVs would require $\lceil\dfrac{N_{\gamma}^v(1)}{L}\rceil\lambda_mT$ $T_{m}$-length slots. Part 1 of Fig. \ref{Fig: slots} shows an example of the slot utilization in a TDD MIMO scenario for one group of $L$ AVs when $\lambda_m=2/T_v=2/T$ and $T_m = T_{\text{slot}}$, where, within duration $T$, there are $\lambda_mT=2$ downlink messages and $LT/T_v=L$ uplink pilots.

Compared with former generations of cellular systems, in 5G, more flexible TDM is allowed,  as shown in Fig. 2 in \cite{lien2017}. Specifically, adjacent slots can be independently used for uplink and downlink transmissions. With such flexibility, the numbers of uplink slots and downlink slots can be arbitrarily specified. %uplink pilots for one group of $L$ AVs can be split into smaller pieces for transmission, such that the small time slot gaps  between messages can be exploited instead of being wasted. As shown in Fig. \ref{Fig: slots}, after adding the messages of the third group of $L$ AVs, the time slot gaps between existing messages can be further filled by the pieces split from the pilots (in pink). 
As shown in Fig. \ref{Fig: slots}, with the increase in the number of AVs, eventually, it reaches a point where the time slots in $T$ are compactly filled by pilots and messages of $3L$ AVs, and no additional AVs can be accomodated. Thus, with flexible TDM in 5G, the sum lengths of all uplink pilots and downlink messages (both in sec) within duration of $T$, can occupy a duration close to $T$. 

For the TDD MIMO scenario in this subsection, let \begin{align}\label{eq:gL}
\lceil\dfrac{N_{\gamma}^v(1)(T/T_v)T_{\text{pilot}}}{T_{\text{slot}}}\rceil T_{\text{slot}}+\lceil\dfrac{N_{\gamma}^v(1)}{L}\rceil(\lambda_mT)T_{m} \leq T
\end{align}
%\begin{align}
%N_{\gamma}^v(1)f_vT_{\text{pilot}}+\dfrac{N_{\gamma}^v(1)}{L}\lambda_mT_{\text{pilot}}\leq 1.
%\end{align} 
%\begin{align}\label{eq:lL}
%N_{\gamma}^v(1)(T/T_v)T_{\text{pilot}}+(\lambda_mT)T_{m} \leq T.
%\end{align}
%As a special case, when $N_{\gamma}^{v_1}(1) =1$, from (\ref{eq:lL}), $v_1\leq\dfrac{1-\lambda_mT_{m}}{f_0T_{\text{pilot}}}v_0$, then, $v_m=\min\{v_l,v_1\}.$
When $T_{\text{slot}}$ is small, $\lceil\dfrac{N_{\gamma}^v(1)(T/T_v)T_{\text{pilot}}}{T_{\text{slot}}}\rceil T_{\text{slot}}\approx N_{\gamma}^v(1)(T/T_v)T_{\text{pilot}}$. Then, we let
\begin{align}\label{eq:gL22}
N_{\gamma}^v(1)(T_{\text{pilot}}/T_v)+\lceil\dfrac{N_{\gamma}^v(1)}{L}\rceil\lambda_mT_{m} \leq 1
\end{align}
Also, when $N_{\gamma}^v(1)=iL$, $\lceil\dfrac{N_{\gamma}^v(1)}{L}\rceil=\dfrac{N_{\gamma}^v(1)}{L}, \forall i\in\mathbb{N}_0$, where $\mathbb{N}_0$ is the set of non-negative integers. And when $jL<N_{\gamma}^v(1)\leq(j+1)L$, $\lceil\dfrac {N_{\gamma}^v(1)}{L}\rceil = j+1,\forall j\in\mathbb{N}_0$. Because $N_{\gamma}^v(1)$ is an integer, the theoretical maximum number of AVs in the cell, from the viewpoint of filling up time slots, is
\begin{align}
\label{eq:gL2}
\hspace{-8pt}N_{\gamma}^v(1) = \left\{ \begin{array}{ll} 
   \Big\lfloor\dfrac{1}{T_{\text{pilot}}/T_v+\dfrac{1}{L}\lambda_mT_{m}}\Big\rfloor, & \text{for}~ N_{\gamma}^v(1)=iL \\
\\
    \Big\lfloor\dfrac{1-(j+1)\lambda_mT_{m}}{T_{\text{pilot}}}T_v\Big\rfloor, &  \text{for}~ jL<N_{\gamma}^v(1) \\ &~~~~~\leq(j+1)L 
 \end{array} \right.
%,
\end{align}
%where the equality is obtained when $N_{\gamma}^v(1)$ is an integer times $L$. %Due to the existence of various communication overhead, the real number of AVs might be smaller than this. 
%Substitute (\ref{eq:gL2}) into Eq. (\ref{eq:flow}),
%\begin{align}\label{eq:flow_final}
%\mathcal{F}_m=\dfrac{\mathcal{L}\mathcal{W}}{\mathcal{C}}\Big\lfloor\frac{1}{\frac{T_{\text{pilot}}}{T_v}+\dfrac{1}{L}\lambda_mT_{m}}\Big\rfloor v=\dfrac{\mathcal{L}\mathcal{W}}{\mathcal{C}}\Big\lfloor\frac{1}{\dfrac{\alpha v f_c}{c}T_{\text{pilot}}+\dfrac{1}{L}\lambda_mT_{m}}\Big\rfloor v
%\end{align}
Given (\ref{eq:gL2}), it can be proved that $\mathcal{F}_m$ in (\ref{eq:flow}) reaches its peak when $N_{\gamma}^v(1) =       \dfrac{1}{T_{\text{pilot}}/T_v+\dfrac{1}{L}\lambda_mT_{m}} = L$; solving this gives $v=\dfrac{c(1-\lambda_mT_{m})}{\alpha f_cT_{\text{pilot}}L}.$ Therefore, the best strategy to maximize $\mathcal{F}_m$ is to choose speed \begin{align}v^*=\min\{v_l, \dfrac{c(1-\lambda_mT_{m})}{\alpha f_cT_{\text{pilot}}L}\},
\end{align}
where $v_l$ is the speed limit imposed by the transport authority. The corresponding theoretical maximum cell sum traffic flow is 
\begin{align}
    \mathcal{F}_m^*=\dfrac{\mathcal{L}\mathcal{W}}{\mathcal{C}}N_{\gamma}^{v^*}(1)v^{*}=\dfrac{\mathcal{L}\mathcal{W}}{\mathcal{C}}\Big\lfloor\dfrac{1}{T_{\text{pilot}}/T_{v^*}+\dfrac{1}{L}\lambda_mT_{m}}\Big\rfloor v^{*}
    \end{align} %It can further validated that when $N_{\gamma}^v(1)$ is not an integer times $L$, the corresponding $\mathcal{F}_m$ is no larger than $\mathcal{F}_m^*,$ so 
%\end{subequations}

%\begin{subequations}
%Similarly, from (\ref{eq:lL}), when $N_{\gamma}^v(1)\leq L$, theoretical maximum number of AVs
%\begin{align}\label{eq:lL2}
%N_{\gamma}^v(1) =       \Big\lfloor\dfrac{1-\lambda_mT_{m}}{T_{\text{pilot}}/T_v}\Big\rfloor.
%\end{align%}
%Substitue (\ref{eq:lL2}) into Eq. (\ref{eq:flow}),
%\begin{align}\label{eq:flow_final2}
%\mathcal{F}_m=\dfrac{\mathcal{L}\mathcal{W}}{\mathcal{C}}\Big\lfloor\dfrac{1-\lambda_mT_{m}}{T_{\text{pilot}}/T_v}\Big\rfloor v
%\end{align%}
%It can be proved that (\ref{eq:flow_final2}) achieves its maximum value whenever $\dfrac{1-\lambda_mT_{m}}{T_{\text{pilot}}/T_v}=i$, where $i$ is an integer smaller than or equal to $L$. Then, the best strategy when $N_{\gamma}^v(1)\leq L$ is to choose speed $v_i^{**}=\min\{v_l, \dfrac{c(1-\lambda_mT_{m})}{\alpha f_cT_{\text{pilot}}i}\} ~\forall~i\leq L$, and the corresponding theoretical maximum sum traffic flow is actually $\mathcal{F}_m^{**}=\mathcal{F}_m^{*}$.

%From the discussion above, we conclude that, for $\mathcal{B}$ frequency channels, maintaining $\mathcal{B}\dfrac{\mathcal{L}\mathcal{W}}{\mathcal{C}}N_{\gamma}^{v^*}(1)v^{*}$ AVs in the cell and let them move at speed $v^*$ is the optimal strategy to achieve the maximum cell sum traffic flow. %Depending on the value of $v_l$, AVs moving at speed values of $v_i^{**}$ might also be optimal strategies.
%Compared with reference speed $v_0$, there is a road-network throughput gain of $\dfrac{N_{\gamma}^{v^*}(1)v^*}{N_{\gamma}^{v_0}(1)v_0}-1$ in the cell.

%\end{subequations}

\begin{figure}     \includegraphics[width=1\linewidth]{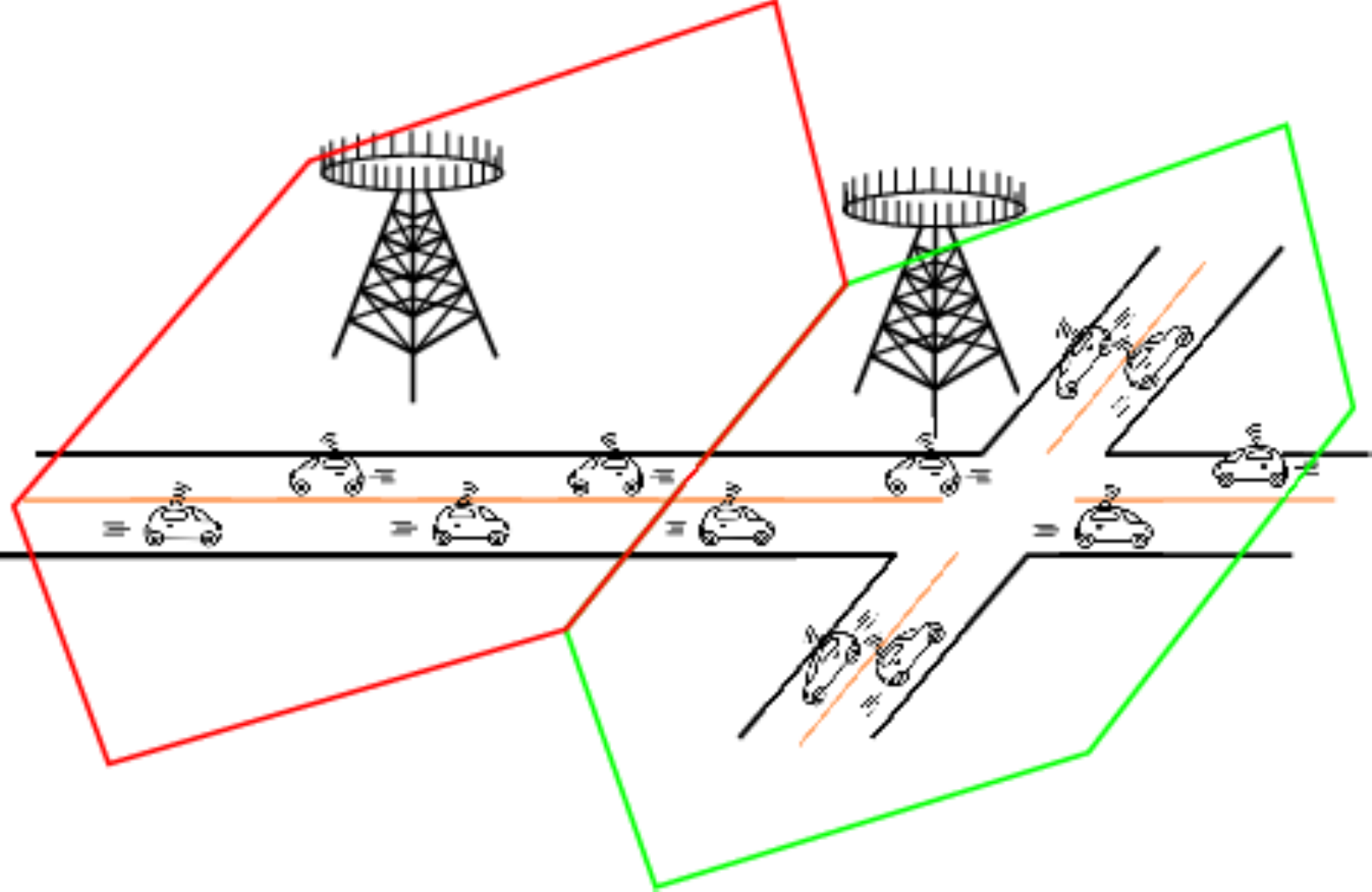}
 \caption{Matched traffic flow across cells}
\label{traffic2}
\end{figure}

\subsubsection{Example in a two-cell scenario}
To demonstrate \emph{Lemma 2}, we consider a frequency channel allocation problem using the two-cell scenario. As shown in Fig. \ref{traffic2}, there are two horizontal lanes in the red cell, and two horizontal lanes and two vertical lanes in the green cell. In the red (green) cell, the AVs move at identical speed and the distances between two adjacent AVs on each lane are identical, thus the lanes' corresponding AV throughputs in the red (green) cell are also identical. This means, for the red cell, the cell sum traffic flow is the same as the sum traffic flow of the cell's horizontal lanes; and for the green cell, the former is twice the latter. Then, to achieve Pareto-optimal road-network throughput, intuitively, the number of frequency channels assigned to the green cell, $\mathcal{B}_g$, should be larger than that of the red, $\mathcal{B}_r$, to make the traffic flows on the horizontal road smooth. $\mathcal{B}_0=\mathcal{B}_g+\mathcal{B}_r$. 

Denote the tuple of the number of lanes, the cell coverage, the optimal AV speed, and the maximum number of AVs, for the green cell, as $(\mathcal{L}_g=4, \mathcal{C}_g, v_g,\mathcal{N}_{g,\Gamma}^{v_g}(\mathcal{B}_g) )$, and for the red cell as $(\mathcal{L}_r=2, \mathcal{C}_r, v_r,\mathcal{N}_{r,\Gamma}^{v_r}(\mathcal{B}_r))$. Then, using Lemma 2, the sum traffic flow of the red cell's horizontal lanes is the same as that of the green cell', and the cell sum traffic flow of the green cell is twice that of the red
\begin{equation}\label{eq:lm2}
 \dfrac{\mathcal{L}_g\mathcal{W}}{\mathcal{C}_g}\mathcal{N}_{g,\Gamma}^{v_g}(\mathcal{B}_g)v_g=2\dfrac{\mathcal{L}_r\mathcal{W}}{\mathcal{C}_r}\mathcal{N}_{r,\Gamma}^{v_r}(\mathcal{B}_r)v_r   
\end{equation}
%, with which the optimal values of $\mathcal{B}_g$ and $\mathcal{B}_r$ can be computed. 

%(THIS FIRST SECTION DOESNT MAKE ENGLISH SENSE) 
For the rest of this section, we consider the special case when the geometries of the two cells are identical; specifically, $\mathcal{C}_g=2\mathcal{C}_r,~ v_g=v_r$, and the two functions  $\mathcal{N}_{g,\Gamma}^{v}(\mathcal{B})=\mathcal{N}_{r,\Gamma}^{v}(\mathcal{B})$. Then, solving Eq. (\ref{eq:lm2}), the optimal number of allocated channels are  $\mathcal{B}_g = 2\mathcal{B}_r = 2/3\mathcal{B}_0$. Compared with equal channel allocation $\mathcal{B}_r=\mathcal{B}_g = 1/2\mathcal{B}_0$, it can be verified that $\mathcal{B}_g = 2\mathcal{B}_r = 2/3\mathcal{B}_0$ brings an additional road-network throughput gain of 33.3\% for this two-cell scenario. %\textbf{$f_D(v) = \dfrac{v}{c}f_c$, where $f_c$ is the carrier frequency and $c$ is the speed of light. }

\subsubsection{Numerical results} 
\begin{figure}          \includegraphics[width=0.9\linewidth]{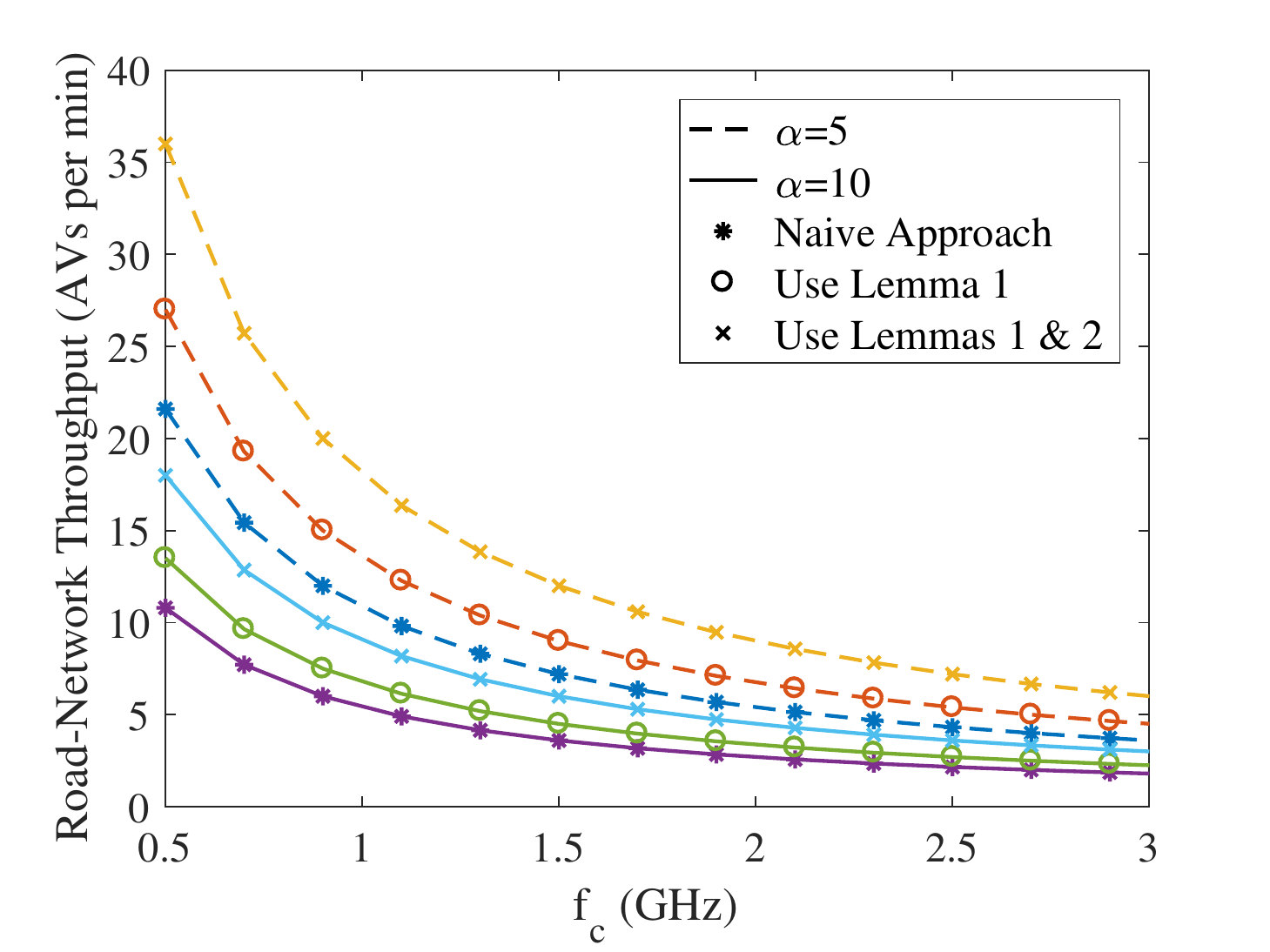}
        %\vspace{-80pt}
                \caption{Road-network thoughput w.r.t carrier frequency $f_c$  ($\lambda_mT_{m}=0.25$).}
           
\label{traffic_020}
\end{figure}

\begin{figure}          \includegraphics[width=0.9\linewidth]{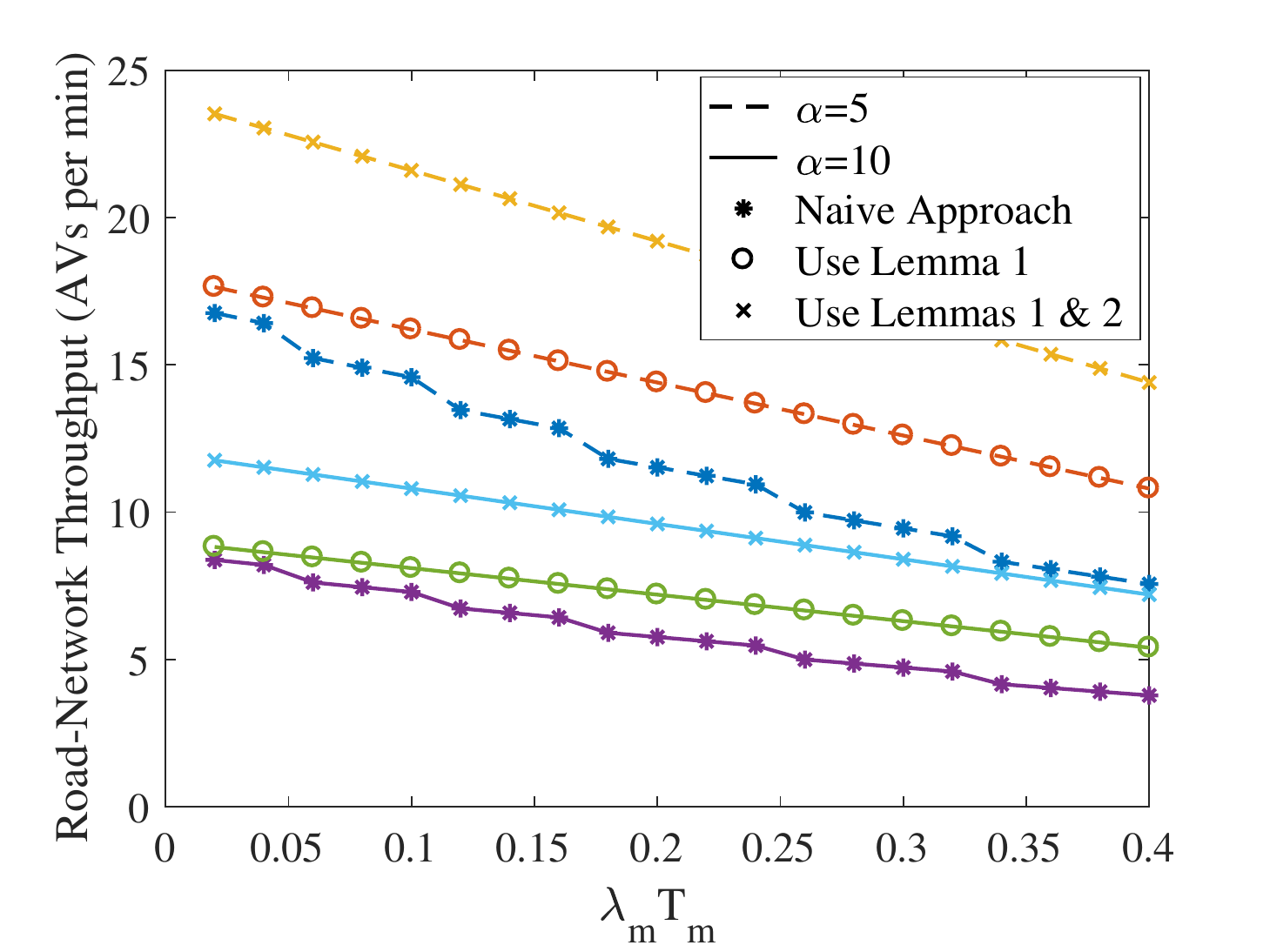}
        %\vspace{-80pt}
                \caption{Road-network thoughput w.r.t $\lambda_mT_{m}$ ($f_c=1$ GHz).}
           
\label{traffic_040}
\end{figure}

Here, we use the one-cell scenario derivations in this subsection for both the red and green cells in Fig. \ref{traffic2}. Let $\mathcal{W}=3$ m and the BS coverage of the red cell $\mathcal{C}_r=12,000$ m$^2$. $L=10$. Since 5G allows mini-slot of duration 0.1 ms, for the length of uplink channel measurement pilot, we let $T_{\text{pilot}}= 0.5$ ms \cite{lien2017}. %and consider a minimum channel measurement frequency $f_0=200$ Hz required for the reference speed $v_0$. 
According to \cite{3gpp2018_2,ge2019}, $\lambda_m\in[10,100]$ Hz and $T_{m}\in[1, 100] $ ms. For simplicity, it is assumed that $v_l$ is large enough to be neglected. Then, for a total of $\mathcal{B}_0=10$ frequency channels, the theoretical road-network throughputs for the scenario in Fig. \ref{traffic2} are plotted  in Figs. \ref{traffic_020} and \ref{traffic_040}, for different values of $\alpha$, $f_c$, $\lambda_mT_{m}$, and different traffic control approaches. ``Naive Approach" is the benchmark when neither of the optimizations in Lemma 1 and 2 are used; specifically, in both cells, AVs move at a suboptimal speed $v^*/2$ and $\mathcal{B}_0$ is equally allocated to the two cells. In contrast, ``Use Lemma 1" uses optimal speed $v^*$ and ``Use Lemma 1 \& 2" utilizes the optimization results of both Lemma 1 and 2. 

From Fig. \ref{traffic_020}, we can see that, the theoretical road-network throughput decreases with $f_c$ and $\alpha$. It means that the throughput decreases with the required amount of channel measurements (the increase of channel measurements is caused either by faster-varying channel as indicated by $f_c$, or by more stringent channel accuracy requirement as indicated by $\alpha$). From Fig. \ref{traffic_040}, we can see that the theoretical road-network throughput decreases with $\lambda_mT_{m}$. It means that this throughput decreases with the amount of downlink messages required by AVs (the increase of downlink messages is caused either by being sent more frequently as indicated by $\lambda_m$ or by longer messages as indicated by $T_{m}$). The unsmoothness of ``Naive Approach" curves in Fig. \ref{traffic_040} is due to the floor operator in Eq. (\ref{eq:gL2}).% Also, it can be validated that, compared with ``Use Lemma 1," ``Use Lemma 1 \& 2" has a throughput gain of 33.3\%.

%\begin{figure}           \includegraphics[ width=1\linewidth]{cdf_time.pdf}
        %\vspace{-80pt}
                %\caption{add a result figure here.}
           
%\label{cdf_time}
%\end{figure}

% to make the traffic flow on the horizontal road smooth, so that optimal overall road-network throughput can be achieved.

% flow optimization: each source destination pair would have multiple good choices using Dijkstra algorithm. Then variabl

% Fluidly like routing is the most optimal. (Give an explanation why) 

% The vision brought is that, there will be no lane in the future maybe. The road is any area or cellular network to make use of all the possible spaces between buildings to pavements or even spaces instead of building straight road. This can achieve larger throughput than any current traffic management, and is the goal of future transportation. How to approach this limit step by step is an important issue to be discussed. This can be more exciting for UAVs, as there might be more traffic for UAVs, it would be like a high way in the air without any concrete and steel, totally made by cellular signals. 

\section{Conclusion and Future work}\label{last}

%Communication has become an important component of daily life, with good quality of which, AVs can provide passengers more applications for work and entertainment. 
In this paper, the problems of CCR and CCTC are motivated and investigated.  %A vision is provided that AV routing and AV communication can be jointly planned to provide a better overall driving experience. 
For CCR, a two-layered routing scheme, including both inter-$\gamma$-rate-cell and  intra-$\gamma$-rate-cell routing, is proposed, and its performance is compared with two greedy solutions. For CCTC, optimal effective speed and optimal frequency channel allocation across adjacent cells are derived to maximize theoretical road-network throughput within each cell and Pareto-optimal road-network throughputs across multiple cells. %This concept will be useful in future intelligent transportations with communication constraints.

The CCR and CCTC problems have variations that remain to be formulated and resolved. For instance, in CCR, except rate requirement, AV applications may have various communication QoS and edge computing requirements \cite{qualcomn2018}. The co-existence of AVs and other cellular users can change the communication resources in each cell available for AVs, in which case the graph of road network is time-varying \cite{huang2012}-\cite{huang2016}. For CCTC, except maximizing road-network throughput, there are other goals in AV traffic control, for instance, minimizing trip duration and increasing energy efficiency. %Furthermore, in this paper mostly vehicle-to-infrastructure communication is considered; however, vehicle-to-vehicle communication can also play an important part in helping AVs platooning and making joint routing decisions cooperatively. %Furthermore, for larger road-networks, for example, some cross-country networks, less complex routing and preprocessing methods may be required for the sake of practical application. 
Also, in practice, more advanced technologies, like optimal subchannel allocation for broadband communication and the collaboration of multiple BSs, can be used to increase
 the number of AVs one BS (or BSs) control simultaneously. In addition, some practical factors might be involved for more accurate modeling in CCR and CCTC, for instane,  acceleration (deceleration) and group behavior in AV movement control; different AVs may have heterogeneous communication requirements and LoAs, and different BSs may provide heterogeneous communication capabilities; and the road network may be complex involving  road signs, lane merging, roundabout, and parking capability. %, it mentions that communication service can be reserved beforehand. Then, online routing algorithms should update the graph as frequent as possible \cite{huang2012}-\cite{huang2016}. For example, for a certain cell, during a certain time period, if it is already known that there would be many AVs inside and few communication resources left, to navigate a newly accessed AV, all the edges on the graph that move into this cell should be removed for this time-period. 
%(how accurate the ERM and ESM can be built and their impact)

%We have proposed routing approaches in the sections before. But we need to change the  graph to eliminate those edges such that the edges are time-varying. That is, 

% ``A AV-to-infrastructure communication based
% algorithm for urban traffic control
% ''

% Each lane has different velocity flow, which affects the speed of the AVs, and in turn affect the effect of communication

% Arrange the AVs such that the ones who communicate a lot are apart from each other such that the interference is smaller with each other.

% AVs platooning, especially trucks, can save energy.

% You can check the work of 
% Marco Pavone for the routing of robots.

% some interesting literatures
% Using a Distributed Roadside Unit for the Data Dissemination Protocol in VANET With the Named Data Architecture: the BSs are 1.5 km apart

%: DEEP: Density-Aware Emergency Message Extension Protocol for VANETs

% Road Side Unit Deployment: A Density-Based Approach
% V2I can rebroadcast the messages

% \subsection{some related works}

% http://ieeecss.org/sites/ieeecss.org/files/documents/IoCT-Part4-13VehicleToVehicle-HR.pdf

% $http://www.sfbayite.org/wp-content/uploads/2014/07/Presentation-March-2014_Maile.pdf$

% https://automotivelectronics.com/vehicle-to-infrastructure-communication/
\end{document}